\documentclass[a4paper,fleqn]{cas-dc}
\usepackage{amssymb}
\usepackage{amsmath}
\usepackage{graphicx}
\usepackage{booktabs}
\usepackage{multirow}
\usepackage{makecell}
\usepackage{array}
\usepackage{tabularray}

\usepackage{caption}
\usepackage{subcaption}
\usepackage{anyfontsize}
\usepackage{url}
\usepackage{pifont}
\usepackage{lscape}
\usepackage{acro}
\usepackage{enumitem}
\usepackage{wasysym}
\usepackage[numbers, sort]{natbib}
\usepackage{multirow}
\usepackage{tabularx}

\usepackage{listings}
\lstdefinestyle{bashstyle}{
    language=bash,
    basicstyle=\footnotesize\ttfamily,
    frame=single,
    breaklines=true,
    showstringspaces=false,
    numbers=none,
    keywordstyle=\color{red},
    commentstyle=\color{gray},
    stringstyle=\color{teal}
}

\lstdefinestyle{jsonstyle}{
    language=,
    basicstyle=\footnotesize\ttfamily,
    frame=single,
    breaklines=true,
    showstringspaces=false
}

\usepackage{tikz}
\usetikzlibrary{matrix}

\usepackage[many]{tcolorbox}    	
\usepackage{lipsum}

\tcbset{
    sharp corners,
    colback = white,
    before skip = 0.2cm,    
    after skip = 0.5cm      
}                           

\newtcolorbox{mybox}{colback=gray!5!white, colframe=gray!5!white}
\newtcolorbox{boxK}{
    sharpish corners, 
    boxrule = 0pt,
    toprule = 4.5pt, 
    enhanced,
    fuzzy shadow = {0pt}{-2pt}{-0.5pt}{0.5pt}{black!35}, 
    fontupper = \sffamily
}

\definecolor{sub}{HTML}{F5F5DC} 
\newtcolorbox{boxC}{
    colback = sub, 
    boxrule = 0.25pt,  
    breakable
}

\definecolor{subC2}{HTML}{F0F8FF} 
\newtcolorbox{boxC2}{
    colback = subC2, 
    boxrule = 0.25pt,  
    breakable
}

\usepackage{tikz}

\usepackage[utf8]{inputenc}

\newcommand{\partially}{\(\sim\)} 
\newcommand{\totally}{\ding{51}}  
\newcommand{\notatall}{\ding{55}} 

\begin{document}
\let\WriteBookmarks\relax
\def\floatpagepagefraction{1}
\def\textpagefraction{.001}

\shorttitle{TwinArch: A Digital Twin Reference Architecture}

\shortauthors{Somma \textit{et al.}}  
\title[mode=title]{TwinArch: A Digital Twin Reference Architecture}

\author[1]{Alessandra Somma}
\ead{alessandra.somma@unina.it}
\affiliation[1]{organization={University of Naples Federico II},
            city={Naples},
            postcode={80125}, 
            country={Italy}}

\author[1]{Domenico Amalfitano}
\ead{domenico.amalfitano@unina.it}

\author[1]{Alessandra {De Benedictis}}
\ead{alessandra.debenedictis@unina.it}

\author[2]{Patrizio Pelliccione}
\ead{patrizio.pelliccione@gssi.it}
\affiliation[2]{organization={Gran Sasso Science Institute (GSSI)},
            city={L'Aquila},
            postcode={67100}, 
            country={Italy}}

\begin{abstract}
\noindent \textit{\textbf{Background.}} Digital Twins (DTs) are dynamic virtual representations of physical systems, enabled by seamless, bidirectional communication between the physical and digital realms. 
Among the challenges impeding the widespread adoption of DTs is the absence of a universally accepted definition and a standardized DT Reference Architecture (RA). 
Existing state-of-the-art architectures remain largely domain-specific, primarily emphasizing aspects like modeling and simulation. Furthermore, they often combine structural and dynamic elements into unified, all-in-one diagrams, which adds to the ambiguity and confusion surrounding the concept of Digital Twins. 

\noindent \textit{\textbf{Objective.}} To address these challenges, this work aims to contribute a domain-independent, multi-view \textit{Digital Twin Reference Architecture} that can help practitioners in architecting and engineering their DTs. 

\noindent \textit{\textbf{Method.}} 
We adopted the \textit{design science} methodology, structured into three cycles: \textit{(i)} an initial investigation conducting a Systematic Literature Review to identify key architectural elements, \textit{(ii)} preliminary design refined via feedback from practitioners, and \textit{(iii)} final artifact development, integrating knowledge from widely adopted DT development platforms and validated through an expert survey of 20 participants.


\noindent\textit{\textbf{Results.}} The proposed Digital Twin Reference Architecture is named \textbf{TwinArch}. 
It is documented using the \textit{Views and Beyond} methodology by the Software Engineering Institute.  TwinArch website and replication package: \url{https://alessandrasomma28.github.io/twinarch/}.

\noindent
\textit{\textbf{Conclusion.}} TwinArch offers practitioners practical artifacts that can be utilized for designing and developing new DT systems across various domains. It enables customization and tailoring to specific use cases while also supporting the documentation of existing DT systems.

\end{abstract}

\begin{highlights}
    \item Existing DT architectures are domain-specific, they are not documented with multi-views, and merge in the same view structural and dynamic elements.
    \item Practitioners reported challenges in applying current DT standards.
    \item TwinArch integrates literature elements, insights from practitioners and DT development platforms. 
    \item TwinArch is documented using the Views and Beyond method aligned with ISO 42010.
    \item Completeness, usefulness, usability of TwinArch are validated by DT experts.
\end{highlights}

\maketitle



\section{Introduction}
\label{sec:introduction}
\textbf{Digital Twins} (DTs) are dynamic virtual representations of physical systems, enabled by seamless, bidirectional communication between the physical and digital realms \cite{RW_CharacterisingDT20, RW_DTARchetypes}. Unlike traditional simulators, DTs are continuously updated with real-world data, supporting advanced functionalities like predictive maintenance, real-time monitoring, and system control \cite{tao_dtindustry, dt_barricelli19}. These capabilities have positioned DTs as a transformative technology finding applications across diverse fields such as manufacturing, aerospace, and automotive industries \cite{dt_barricelli19, RW_SurveyDT24}.

Although the concept of Digital Twins has existed for nearly two decades, interest from both academia and industry has surged only recently, and a universally accepted definition has yet to be established \cite{RW_DTArchitecturalModels}. Tao \textit{et al.} \cite{tao_dtenabling} introduced a five-dimensional model for digital twins, defined as:
$DT = \{P_s, V_s, DD, S_s, CN\}$, where $P_s$ represents the physical space of real-world entities and interactions, $V_s$ denotes the virtual space with digital replicas dynamically reflecting physical behavior, $DD$ comprises real-time and historical data enriched with domain knowledge, $S_s$ provides services such as monitoring and prediction \cite{M35_ProcessPred21}, and $CN$ ensures seamless integration across these dimensions.

Despite the growing adoption and potential benefits of Digital Twins, both academia and industry face substantial challenges in unlocking their full potential. The inherent complexity of DT systems, coupled with high-development costs and time-intensive maintenance, remains a major barrier to broader adoption \cite{RW_SurveyDT24, RW_SurveyDTs, M06_DTPDM24}. A significant contributing factor to these challenges is the absence of a software \textbf{Reference Architecture} (RA) to systematically guide the design, development, and maintenance of these software-intensive systems, regardless of their application domain \cite{RW_SurveyDT24, RW_SurveyDT22, DTFrameworksSurvey24}.

A RA provides an abstraction of software components, their roles, and their interactions, serving as a template or blueprint for creating concrete software architectures. It encapsulates the essential characteristics of systems within a particular domain and offers a structured framework for designing new systems or enhancing existing ones \cite{SoftwareRAJSS}. The importance of RAs in software development, particularly for DT systems, is evidenced by initiatives led by standardization bodies. For instance, ISO is developing: \textit{(i)} a standard for RAs in enterprises, systems, and software (ISO/IEC/IEEE CD 42042\footnote{ISO/IEC/IEEE CD 42042 standard, \textit{“Enterprise, systems and software — Reference architectures”}, \url{https://www.iso.org/standard/87310.html}}), and \textit{(ii)} a standard for a DT Reference Architecture (ISO/IEC AWI 30188\footnote{ISO/IEC AWI 30188 standard, ``\textit{Digital Twin — Reference architecture}'', \url{https://www.iso.org/standard/53308.html}}), which are both still in the early stages of development. 

Indeed, the standardization process is lengthy and involves multiple stages, often spanning several years. Moreover, the complexity of standards and the absence of practical tools to facilitate their adoption often hinder their implementation in real-world projects. For example, the ISO 23247 standard\footnote{ISO 23247 standard, \textit{“Automation systems and integration — Digital twin framework for manufacturing”}, available at: \url{https://www.iso.org/standard/75066.html}}, which offers a reference architecture for Digital Twins in manufacturing \cite{ISO23247}, has faced criticism for its limited applicability due to its domain-specific nature \cite{ISO23247AEROSPACE, DTBatteryISO}. Additionally, it has been noted that the standard lacks critical components, such as those related to data management \cite{EdgeDTISO}. Even within the manufacturing sector, practitioners have reported challenges in applying the standard effectively, as highlighted in \cite{RW_StandardizationDTArch}, largely due to the absence of supporting artifacts for DT instantiation.

Given the limitations and slow adoption of standardization efforts, the DT research community has proposed various reference and software architectures to address this gap. However, most of these proposals are either domain-specific or tailored to particular services~\cite{M33_SelfAdaptiveDT21, M20_ArchitectingDTDD23, M16_ArchitecturePdm23, M11_DTAnomalyTII23}, which constrains their flexibility and reduces their broader applicability and usefulness across various DT domains. Additionally, many of these architectures rely on a single, unified diagram combining structural elements from different abstraction levels, often overlooking dynamic aspects~\cite{M09_maketwin24}. These all-in-one design approaches contradict the ISO/IEC/IEEE 42010 standard\footnote{ISO/IEC/IEEE 42010:2022 standard, \textit{“Software, systems and enterprise — Architecture description”}, available at: \url{https://www.iso.org/standard/74393.html}}, which recommends to document architectures by using 
multiple \textit{architectural views}. 

Architectural views highlight specific subsets of system elements and their relationships, designed to address the concerns of particular stakeholders \cite{ISO42010}. However, even when attempts are made to separate architectural concerns into distinct views, the resulting DT architectures lack a holistic perspective, instead focusing narrowly on specific aspects of Digital Twins, such as modeling and simulation capabilities \cite{M41_DTFourLayer}. This limitation, influenced in part by the subjective interpretation of what defines a DT, results in fragmented architectures that fail to integrate structural and dynamic elements comprehensively. Consequently, existing solutions tend to be highly customized and tailored to specific cases, highlighting the pressing need for a multi-view and domain-independent Reference Architecture for DTs.

The {\bf goal} of this work is to identify a domain-independent, multi-view Digital Twin Reference Architecture that can help practitioners in architecting and engineering their DTs.
The proposed \textit{Digital \textbf{Twin} Reference \textbf{Arch}itecture} is called 
\textbf{TwinArch}, 
and it is organized in multiple views~\cite{ISO42010}, following the \textit{Views and Beyond} (V\&B) methodology proposed by the Software Engineering Institute (SEI). 
TwinArch synthesizes architectural elements from existing DT architectures, integrates feedback from DT practitioners and incorporates insights from three widely adopted DT development platforms—\textit{Eclipse Ditto}, \textit{Azure Digital Twins} (ADT), and \textit{FIWARE}.

TwinArch is designed to deliver scientifically robust and practical artifacts for researchers and practitioners engaged in the design and development of DT systems across diverse domains. These artifacts serve a \textit{dual purpose}: supporting the documentation of existing DT systems and guiding the creation of new ones. Practitioners can utilize TwinArch as a foundational framework, adapting and customizing it to meet the specific requirements of their use cases, while also leveraging the provided mapping between architectural elements and the software tools of the selected platforms for practical implementation.

To perform the study of this paper, 
we employed the \textit{design science} methodology, an iterative and structured approach conducted over three cycles. In the first cycle, a Systematic Literature Review (SLR) was performed to analyze existing DT architectures. The second cycle involved developing an initial draft of TwinArch, which was then preliminarily validated by practitioners participating in the project supporting this work. 
In the third cycle, TwinArch was further refined by integrating knowledge from the three selected platforms, identified through a specific search for DT platform solutions. 

TwinArch's completeness, usefulness, and perceived usability were evaluated through an online survey conducted with DT experts. The results indicated a broadly positive perception of TwinArch, with respondents affirming its completeness, usefulness, and usability. TwinArch was particularly praised for providing clear guidelines and facilitating communication among stakeholders, developers, and researchers. Statistical and practical significance tests further confirmed that TwinArch is well-suited to serve as a complete, useful and usable Reference Architecture for Digital Twins.

The reminder of this paper is organized as follows. Section \ref{sec:related} summarizes related works on DT architectures. Section \ref{sec:methodology} describes the design science methodology, while Section \ref{sec:twinarch} presents the proposed TwinArch. Section \ref{sec:survey} details the online survey results. Section \ref{sec:discussion} discusses challenges in Digital Twin architectures. Finally, Section \ref{sec:conclusion} draws our conclusions and introduces future work.

\section{Related Work}
\label{sec:related}

The growing interest in Digital Twin technology has led to significant efforts to define software architectures that support their design, development, and deployment, as demonstrated by the increasing body of research on the topic~\cite{RW_ArchitectingDT, RW_SurveyDT24, RW_SurveyDTs}. For instance, Bolender \textit{et al.} \cite{M33_SelfAdaptiveDT21} developed a model-driven DT architecture for the self-adaptive manufacturing by incorporating domain-specific modeling and case-based reasoning. Similarly, Maceas \textit{et al.} \cite{M20_ArchitectingDTDD23} employed a domain-driven design methodology to structure DT systems in highly evolving environments. Boyes \textit{et al.} \cite{M21_AnalysisFramework22} proposed an analysis framework to identify common functional characteristics of DTs, addressing ambiguities caused by varying DT definitions. 

Layered patterns are widely used to architect DTs. For example, Redelinghuys \textit{et al.}~\cite{M38_DTSixLayer20} introduced a six-layer architecture to facilitate seamless data and information exchange between cyberspace and the physical twin, inspired by Cyber-Physical Systems (CPSs). Steinmetz \textit{et al.}~\cite{M30_KeyComponentsDT22} defined key components for DT-based systems with varying levels of granularity, organizing these components into four layers to capture system concerns. Similarly, Malakuti \textit{et al.}~\cite{M41_DTFourLayer} proposed an abstract four-layer architecture pattern to integrate information from diverse sources into DTs.

Despite these contributions, many DT architectures rely on single-view approaches. As highlighted in \cite{M21_AnalysisFramework22}, this often results in confusion surrounding the DT concept and improper use of model elements, where structural and dynamic aspects are mixed instead of being represented in distinct views, as recommended by the ISO 42010 standard~\cite{Book_DocumentingSA, ISO42010}. Although multi-view approaches are less common, notable exceptions exist. For instance, Van Dinter \textit{et al.} \cite{M16_ArchitecturePdm23} proposed a multi-view reference architecture for DT-based predictive maintenance systems, organizing views into user, structural, and layered perspectives. Similarly, Tao \textit{et al.} \cite{M09_maketwin24} introduced the makeTwin architecture, specifying ten functional modules for rapid DT prototyping and deployment, instantiated in the manufacturing domain. 

The ISO 23247 standard, published in 2021 \cite{ISO23247}, defines a DT reference architecture specifically for manufacturing. It includes an entity-based reference model with four main entities: \textit{(i)} Device Communication (responsible for data collection and control of Observable Manufacturing Elements, such as physical assets), \textit{(ii)} Digital Twin (focused on modeling, synchronization, and management), \textit{(iii)} User (hosting applications that utilize DT services), and \textit{(iv)} Cross-System (providing overarching functionalities like security and data translation)~\cite{ISO23247Analysis}. The standard also includes a functional view, which details Functional Entities (FEs) to specify the functionalities at each level of the reference model.

Several studies have adopted ISO 23247 to design DT software architectures. For example, Bong Kim \textit{et al.}~\cite{ManufacturingDTISO} proposed a DT architecture for additive manufacturing to address process variability and enhance quality assurance. Spaney \textit{et al.}~\cite{ModelDrivenDTISO} presented a standard-based model-driven DT architecture for milling processes, Melo \textit{et al.}~\cite{DTAssemblyLineISO} applied the ISO 23247 standard to develop a DT for automotive assembly lines, focusing on process precision. Wallner \textit{et al.}~\cite{DTManufacturingCellISO} extended the standard to design a DT for flexible manufacturing cells, integrating lifecycle management, path planning, and collision detection to manage reconfigurations. Caiza \textit{et al.}~\cite{DTFlexibleManufacturingISO} implemented an immersive DT architecture using augmented reality for real-time monitoring and control.

While these studies demonstrate the applicability of ISO 23247 across manufacturing scenarios, challenges remain in extending its use to other domains. For instance, Ferko \textit{et al.}~\cite{DTBatteryISO} explored its application in battery systems. Similarly, Shtofenmakher \textit{et al.}~\cite{ISO23247AEROSPACE} attempted to tailor the standard for aerospace use, focusing on on-orbit collision avoidance. Despite these efforts, researchers have highlighted significant limitations in ISO 23247, including domain-specific constraints, perceived misalignments and lack of concrete tools supporting the instantiation of DTs. A recent industrial survey \cite{RW_StandardizationDTArch} underscored the practitioners' difficulties in adopting the standard. For instance, they noted the absence of important Functional Entities, such as those for data storage and management. This need is further validated by Kang \textit{et al.} \cite{EdgeDTISO}, who extended ISO 23247 with edge computing technologies to enhance data processing and decision-making in DT systems.

In line with the ISO 23247 standard, which is tailored to manufacturing, most of the DT architectures available in the literature are designed for specific domains or services, such as manufacturing~\cite{M33_SelfAdaptiveDT21}, car-as-a-service~\cite{M30_KeyComponentsDT22}, transportation~\cite{M01_BlockchainDT24}, water treatment~\cite{M07_DTWaterPlatform24}, and predictive maintenance \cite{M16_ArchitecturePdm23}. This domain-specific focus limits their flexibility and applicability across diverse contexts, underscoring the need for a domain-independent and multi-view architecture. To address this challenge, the ISO organization is actively working on a DT reference architecture through initiatives like ISO 30188 (under development). 

Summarizing, the discussed challenges 
and limitations of current DT architectures, highlight the need for a domain-independent, multi-view DT reference architecture. This study addresses these issues by: \textit{(i)} documenting TwinArch using the Views and Beyond method in alignment with ISO 42010; \textit{(ii)} integrating state-of-the-art architectural elements into TwinArch, informed by feedback from DT researchers and insights from three well-known DT development platforms; and \textit{(iii)} offering reusable artifacts that can be applied by DT practitioners across multiple domains.

\section{Methodology}
\label{sec:methodology}
\begin{figure}[!h]
\centerline{\includegraphics[width=\columnwidth]{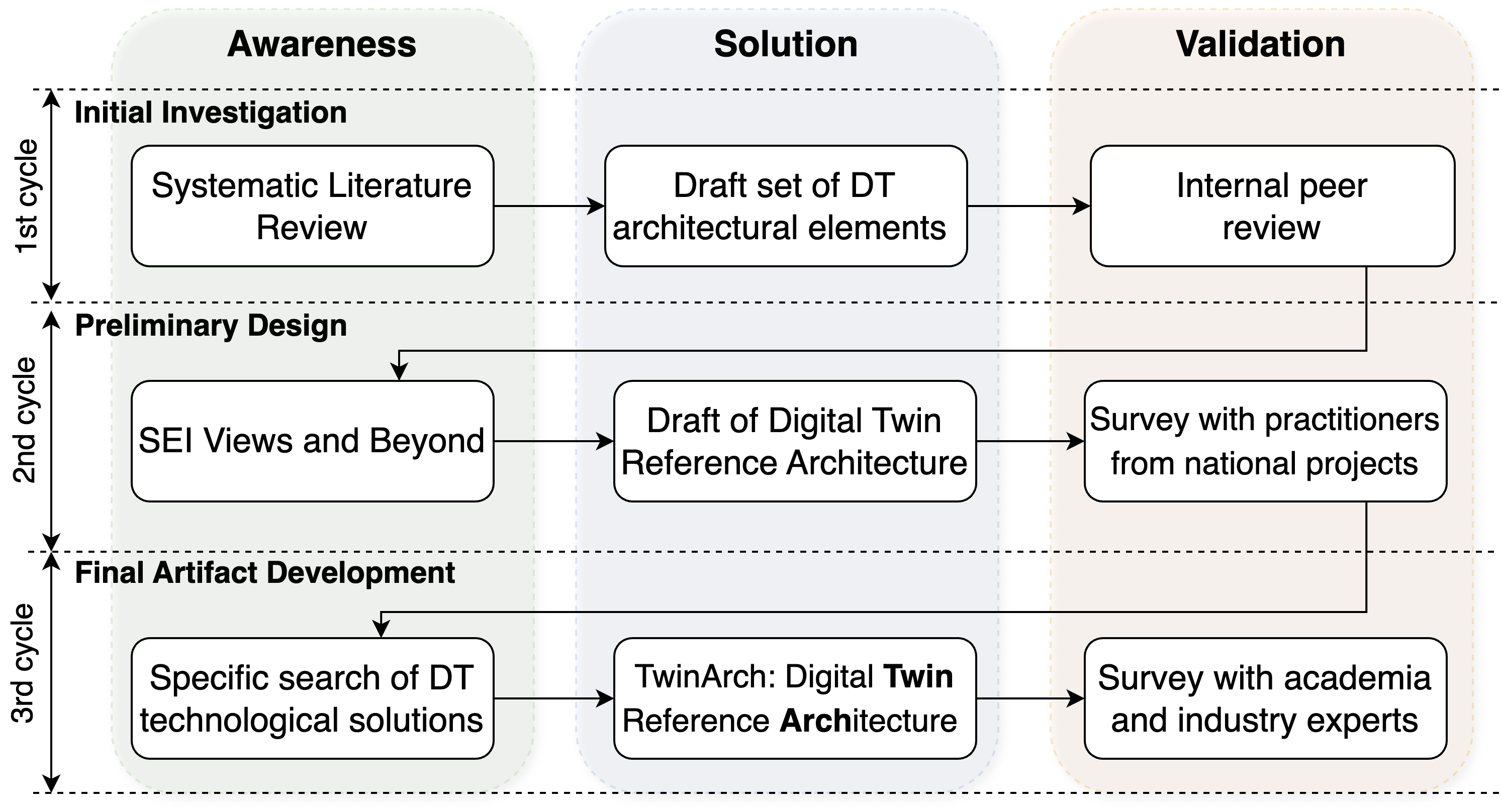}}
\caption{Overview of activities for designing TwinArch using the design science methodology.}
\label{fig:dsrprocess}
\end{figure}

Our work contributes to the definition of \textbf{TwinArch}, the \textit{Digital \textbf{Twin} Reference \textbf{Arch}itecture}. To design TwinArch, we adopted the \textit{design science} methodology, a structured process comprising three primary phases: \textit{context awareness}, \textit{solution synthesis}, and \textit{solution validation} \cite{DSRBook14, DSRWieringa09}. As shown in Figure \ref{fig:dsrprocess}, which illustrates our methodology steps, we adopted an iterative process organized into three cycles, each encompassing the design science phases to understand the problem context, devise a solution, and validate it. 

\textbf{First Cycle: Initial investigation.} The first research cycle aimed to build awareness of the current state-of-the-art in Digital Twin architectures. To achieve this, a Systematic Literature Review was conducted to identify core Digital Twins' architectural elements and uncover limitations in existing approaches, as detailed in Section \ref{sec:discussion}. The identified elements were then refined and validated through internal peer reviews and collaborative discussions with co-authors, resulting in a consensus on the final set of elements and their responsibilities, forming a solid basis for the next research cycle.


\textbf{Second Cycle: Preliminary design.} In the second research cycle, the internally validated architectural elements formed the basis for the initial draft of the Digital Twin Reference Architecture. This draft was designed and documented using the SEI Views and Beyond method, which was analyzed during the context-awareness phase. The TwinArch draft and its elements were subsequently presented to five practitioners participating in one of the national projects supporting this study (see Sec. \ref{sec:ack}).

The preliminary evaluation involved two academic experts and three industry professionals working on Digital Twin case studies, ensuring a balanced assessment by combining theoretical insights with practical experience. Their diverse expertise provided valuable feedback, including recommendations to align the architectural elements to existing Digital Twin technological solutions and to incorporate detailed, practical examples. For instance, practitioners asked questions like, \textit{“How does this align with platforms for DT development?”} and \textit{“Can you provide examples from existing frameworks?”}. These suggestions were integrated into the final iteration.

\textbf{Third Cycle: Final Artifact Development.} The third cycle focused on refining TwinArch by integrating feedback and suggestions received during the previous iteration, and on performing a final validation through the collection of broader feedback from DT experts of academia and industry.

To address comments from practitioners during the previous cycle, a specific search of DT platforms and frameworks was carried out. Among available solutions, we selected Eclipse Ditto, Azure Digital Twins, and FIWARE as the reference platforms for our study due to their popularity in both industrial and research projects.  
This in-depth examination provided a comprehensive understanding of each platform's specific characteristics, enabling us to refine the architectural elements to better align with existing Digital Twin solutions and incorporate detailed, practical examples. 

Finally, TwinArch was validated through an online survey involving 20 Digital Twin experts from both industry and academia to assess its completeness, usefulness, and perceived usability.
These experts were identified through Digital Twin communities on platforms like LinkedIn and X (formerly Twitter), as well as professionals actively involved in DT-related projects. The feedback from the experts proved invaluable, identifying potential areas for future improvement, such as including support for domain-specific instantiation of TwinArch. 

The rest of this Section provides details on the design science process focusing on the most relevant sub-phases. More specifically, Section \ref{sec:firstcycle} covers the initial investigation conducted through the literature review. Section \ref{sec:secondcycle} focuses on the methodology adopted during the preliminary design iteration, centered around the Views and Beyond method. Section \ref{sec:thirdcycle} refers to the final artifact development cycle and details the exploration of Digital Twin platforms and the online survey conducted for final validation. 


\subsection{Initial investigation: Building Awareness with Literature Review}
\label{sec:firstcycle}
\begin{figure*}[!t]
\centerline{\includegraphics[width=0.65\textwidth]{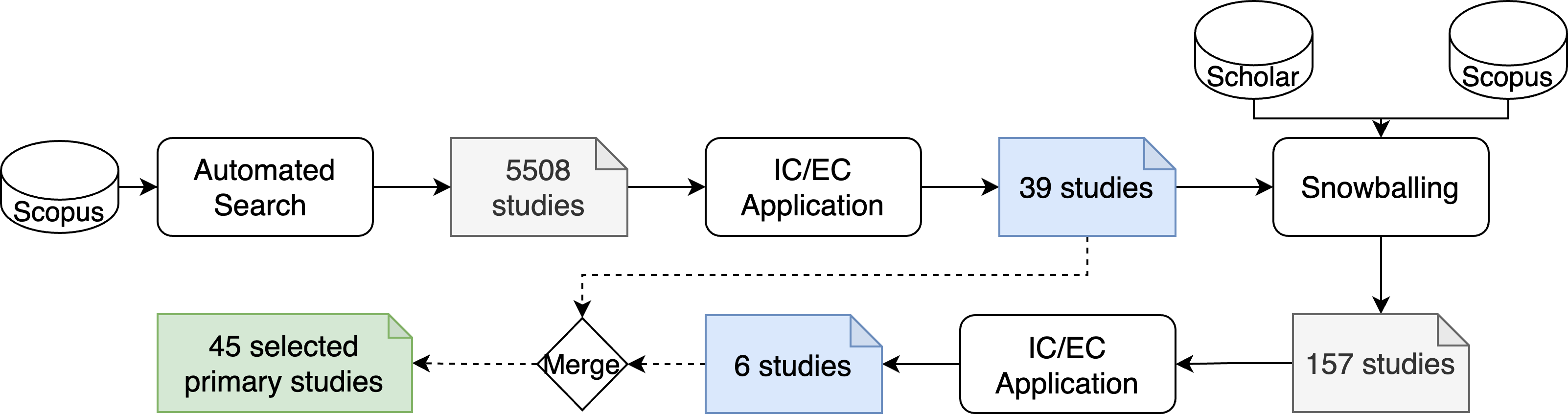}}
\caption{Systematic Literature Review process.}
\label{fig:slrselection}
\end{figure*}

We conducted a Systematic Literature Review based on the guidelines provided by Petersen \textit{et al.} ~\cite{SMSPetersen}. To ensure clarity and rigor, we defined a precise review protocol, defining the research goals, following the structured review process, and extracting data, while implementing measures to mitigate potential threats to the validity of the results. Further details can be found in the replication package.


\noindent \textbf{Review Process.} 
Our literature review began by defining research questions, from which a list of terms, synonyms and abbreviations was compiled. Following the guidelines of Kitchenham and Charters \cite{kitchenham2009systematic}, a search string was constructed using the conjunction (AND) of disjunctions (OR) of the selected terms. The finalized search string is shown in the following box:

{\small \begin{boxK}
\begin{center}
(``Digital Twin'' \textbf{OR} ``Virtual Twin'' \textbf{OR} ``Digital Replica'' \textbf{OR} ``Virtual Replica'') \textbf{AND} (Architect* \textbf{OR} Framework \textbf{OR} Platform \textbf{OR} Document* \textbf{OR} View \textbf{OR} Style)
\end{center}
\end{boxK}
}

Figure \ref{fig:slrselection} illustrates the steps of the selection process, beginning with the execution of the search string in the Scopus database\footnote{The search was conducted in June 2024.}.
The inclusion criteria ensured that only studies directly related to the definition and documentation of Digital Twin architectures, written in English, peer-reviewed, and published in high-ranked journals or conference proceedings, were considered.
The exclusion criteria, on the other hand, filtered out earlier versions of studies, publications conflating Digital Twin concepts with the Metaverse or AI models, and works that treated Digital Twins solely as simulated models. To minimize the risk of overlooking relevant literature, we conducted backward snowballing using Scopus and forward snowballing with Google Scholar\footnote{\url{https://scholar.google.com/}}. This process resulted in a final set of \textit{45 primary studies}.

\noindent \textbf{Data Extraction.} The research goal and the screening phase 
guided the development of a data extraction scheme comprising a set of categories for collecting information from the selected studies.
In addition to capturing the metadata of the publications, the scheme includes elements such as the number and type of architectural views and 
the notations used for documenting these views. 
Using this extraction scheme, data were systematically retrieved by analyzing the selected studies to gather relevant information. 

\noindent \textbf{Results.} The complete SLR process along with the obtained results are reported in the replication package available at \url{https://alessandrasomma28.github.io/twinarch/slr.html}. In summary, the review revealed that only four out of the 45 selected papers presented more than one architectural view, and that the majority of studies relied on informal notations for architecture documentation. Moreover, from the analysis of selected papers we were able to identify some recurrent architectural elements that were used as the basis to build a first draft of the reference architecture (see Table \ref{tab:dtde} and Table \ref{tab:dtc}). 


\subsection{Preliminary Design: Drafting TwinArch with the Views and Beyond}
\label{sec:secondcycle}
The initial draft of TwinArch was designed in the second cycle in accordance with the Views and Beyond method, a widely recognized approach for documenting architectures proposed by the Software Engineering Institute. Unlike fixed-view methods such as the Rational Unified Process, which relies on Krutchen' 4+1 model \cite{krutchen}, the V\&B method prioritizes flexibility, allowing architects to tailor views to the specific concerns and requirements of a given system~\cite{ComparingVBISO}.

The V\&B method organizes the architecture documentation
into three view types \cite{Book_DocumentingSA}. The \textit{Module View} captures the system's decomposition into software modules, each responsible for a cohesive set of functionalities. The \textit{Component-and-Connector (C\&C) View} focuses on the system's runtime structure, representing components (processing units) and connectors (interactions) to highlight operational properties. The \textit{Allocation View} maps the architecture to its physical or organizational environment, illustrating relationships between software and non-software elements, such as hardware or organizational structures.

Each view is defined by an architectural style, 
which specifies the types of elements, their relationships, and constraints on their usage \cite{Book_DocumentingSA, ISO42010}. The method supports various notations for documenting architectures, ranging from informal to semi-formal (e.g., Unified Modeling Language, UML) and formal notations (e.g., ArchiMate). In addition to these core views, the V\&B method incorporates supplementary documentation, referred to as \textit{Beyond} aspects, such as behavioral views to capture dynamic interactions between architectural elements.

In this work, we document TwinArch using the module and component views,
to describe the structural elements of a Digital Twin system. The Allocation View is excluded as TwinArch is a domain-independent reference architecture and does not include deployment details specific to particular application domains \cite{SoftwareRAJSS}. Each view follows the architectural styles recommended by the V\&B method. Furthermore, TwinArch includes behavioral documentation to describe dynamic interactions between elements within each view and traceability across views to ensure coherence in representing the overall Digital Twin system. 

\subsection{Final Artifact Development: TwinArch Refinement and Online Survey}
\label{sec:thirdcycle}
TwinArch draft was refined by incorporating knowledge from three selected Digital Twin development platforms, as outlined in Subsection \ref{sec:dtsol_search}. The finalized TwinArch was validated through an online survey, with the process detailed in Subsection \ref{sec:surveysteps}.

\subsubsection{Specific Search of DT solutions}
\label{sec:dtsol_search}
In the third cycle, a targeted search for Digital Twin solutions was carried out, focusing on practical, open-source or commercial, widely-used platforms and frameworks, i.e. the collections of software tools designed to facilitate the creation, deployment, and maintenance of Digital Twins \cite{ModelingDTPlatforms22}. This search was independently conducted by two of the four authors, with the results and insights thoroughly discussed to reach consensus on refining the architectural elements and finalizing the design of TwinArch.

The search process was carried out using three primary channels: the Google search engine for a broad overview of available platforms; Google Scholar to examine academic discussions and analyses literature on DT technologies; and GitHub repositories and websites to identify open-source projects with active development and community support. A recent survey by Gil \textit{et al.} \cite{DTFrameworksSurvey24} provided a useful starting point by analyzing 14 open-source frameworks and highlighting their varied approaches to offer DT-based services. 

For instance, tools like \textit{Eclipse Ditto} are well-suited for IoT-driven applications, while domain-specific platforms such as the \textit{Digital Twin Cities Centre Platform} (DTCC) focus on smart city planning, and \textit{CPS Twinning} supports cybersecurity focused Digital Twins. These findings emphasize the diversity of DT solutions and their alignment with specific use cases. Building on these insights, additional platforms identified during the search included \textit{Azure Digital Twins} and \textit{FIWARE} \cite{DTPlatforms23, ModelingDTPlatforms22, DTFrameworksSurvey24, M28_DTPlatforms22, FiwarePlatform16}. 
The final selected platforms (Eclipse Ditto, Azure Digital Twins and FIWARE) were chosen for their representation of different categories of Digital Twin solutions, balancing openness, functionality, and domain-specific applicability.


\textbf{Eclipse Ditto}\footnote{\url{https://eclipse.dev/ditto/}} is a free platform developed as part of the Eclipse Internet of Things initiative, aimed at enabling the creation and management of DTs. It abstracts physical devices into faithful digital representations (called \textit{Things}) and provides standardized Application Programming Interfaces (APIs) to allow seamless interaction with these virtual counterparts. 
%
Furthermore, Eclipse Ditto includes a \textit{Gateway} component that facilitates external communication by supporting standard protocols such as MQTT and AMQP, along with the dedicated \textit{Ditto Protocol}. This protocol employs JSON as the message format, enabling Eclipse Ditto to promote interoperability and simplify integration with external systems and services.

\textbf{Azure Digital Twins}\footnote{\url{https://azure.microsoft.com/en-us/products/digital-twins/}} is a cloud-based Platform-as-a-Service solution that provides pay-as-you-go services, enabling the creation of digital models of physical environments. Built on the \textit{Digital Twin Definition Language} (DTDL), a JSON-LD-based schema used for defining Digital Twins, ADT provides a framework for modeling physical entities, their properties, and relationships. Digital Twins are represented as \textit{twin graphs}, dynamic graph-based models that capture the entities and relationships defined using DTDL. 
Moreover, the ADT platform integrates seamlessly with the broader Azure ecosystem. Data from IoT devices are ingested through \textit{Azure IoT Hub}. 
Azure DT also connects with downstream services for analytics, storage, and visualization, such as \textit{Azure Stream Analytics} for telemetry analysis, \textit{Azure Data Lake} for long-term data storage, and \textit{Azure Synapse Analytics} for advanced machine learning workflows. 

\textbf{FIWARE}\footnote{\url{https://www.fiware.org/}} is a free and open-source platform designed to facilitate the development of smart applications across a variety of domains, including smart cities, agriculture, and industry. It offers a modular architecture based on reusable and configurable software components called \textit{Generic Enablers} (GEs), which communicate using the standardized \textit{Next Generation Service Interface} (NGSI) protocol. 
At the heart of every FIWARE-based solution is the \textit{Context Broker}, which manages real-time context data representing the state of physical and digital entities. Moreover, FIWARE offers tools such as \textit{IoT Agents} that facilitate seamless integration with IoT devices by converting native protocols into NGSI format, and FIWARE \textit{Cosmos} for integration with data processing and visualization frameworks. 

FIWARE also spearheads the \textit{Smart Data Models} initiative, which defines domain-agnostic JSON schemas to standardize data structures for smart applications. These models enhance interoperability across systems and platforms, addressing domains such as smart cities, environments, sensors, and agriculture\footnote{\url{https://github.com/smart-data-models/data-models}}. The Digital Twin Definition Language used in Azure is built upon FIWARE data models.

\subsubsection{TwinArch Online Survey}
\label{sec:surveysteps}
TwinArch was validated through an online survey which involved Digital Twin experts recruited from academia and industry. The objective of the survey was to gather practitioner feedback on the three quality factors— completeness, usefulness, and perceived usability— of TwinArch. In line with the Cruzes \textit{et al.} guidelines \cite{thematic}, the online survey conducting process comprised four phases: \textit{(i)} subject selection, \textit{(ii)} questionnaire design, \textit{(iii)} results analysis, and \textit{(iv)} data reporting (see Section \ref{sec:survey}).


\noindent \textbf{Subject Selection.}  TwinArch was created to assist practitioners in designing, developing and documenting DT systems across various domains. To evaluate its completeness, usefulness and perceived usability, survey participants were identified through three main sources: \textit{(i)} authorship and contact details from the papers referenced during the TwinArch design, \textit{(ii)} advertisements on social media and forums, and \textit{(iii)} recognized experts in Digital Twin research. To minimize bias, outreach messages focused on the general objective of developing a reference architecture to support DT system design, without revealing specific details of the proposal. Recruitment took place from September 2024 to January 2025 and concluded when no further responses were received. A total of \textit{20 DT experts} participated in the survey, consisting of 9 industry practitioners and 11 academic researchers. 

\begin{table}[!h]
\centering
\caption{Summary of representative profiles.}
\label{tab:participants}
\resizebox{\columnwidth}{!}{%
\begin{tabular}{|l|l|l|l|l|}
\hline
\textbf{ID} & \textbf{Experience} & \textbf{Affiliation} & \textbf{Role} & \textbf{\# Individuals} \\ \hline
\textbf{P1} & 1 year & Industry & DT Developer & 2 \\ \hline
\textbf{P2} & 1 year & Academia & Assistant Professor & 1 \\ \hline
\textbf{P3} & 1-3 years & Industry & Research Engineer & 2 \\ \hline
\textbf{P4} & 1-3 years & Academia & Researcher/Assistant Professor & 6 \\ \hline
\textbf{P5} & \textgreater{}3 years & Industry & (Senior) Research Engineer & 5 \\ \hline
\textbf{P6} & \textgreater{}3 years & Academia & Associate/Full Professor & 4 \\ \hline
\end{tabular}%
}
\end{table} 

To gather responses, we sent out 546 emails: 367 were directed to DT developers, identified either through works with industrial affiliations or during the specific search for DT platforms, while 179 were sent to paper authors. 
The questionnaire was designed to evaluate the three quality factors for each view. Additionally, respondents were asked two supplementary questions: \textit{(i)} how long they have worked in the DT field, and \textit{(ii)} which DT development platforms they have adopted, if any. 
Based on their answers, we synthesized six representative profiles, as detailed in Table \ref{tab:participants}.

\begin{figure*}[h!]
\centerline{\includegraphics[width=0.65\textwidth]{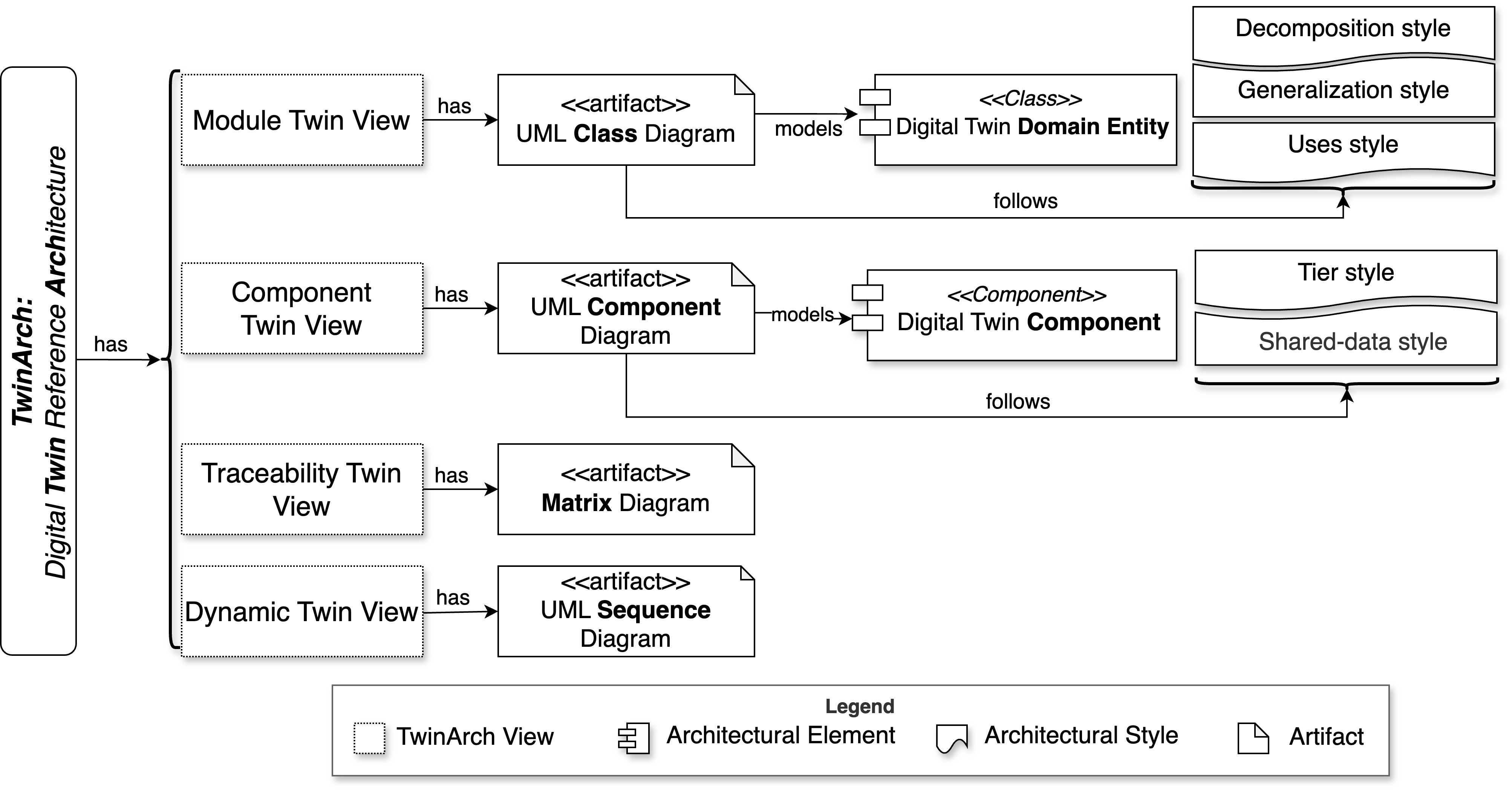}}
\caption{TwinArch Structure.}
\label{fig:twinarchstructure}
\end{figure*}

\noindent \textbf{Questionnaire Design.} The questionnaire was organized into 6 sections, comprising a total of 23 questions with a mix of closed and open-ended formats. The first section gathered background information about the respondents, while sections two through five focused on TwinArch's architectural views. Each of these sections featured a closed question with a rating scale from Strongly Disagree to Strongly Agree to assess three key quality 
factors for each view. The final section assessed the overall TwinArch proposal and included open-ended questions, allowing participants to elaborate on their responses and provide additional insights, including potential strengths and limitations they identified.

\noindent \textbf{Result Analysis.} The data were analyzed using quantitative methods, with statistical tests employed to enhance confidence and provide deeper insights. Specifically, Likert scale responses were visualized using Likert plots and further examined through statistical techniques, including box plots and significance testing. The survey results are detailed in Section~\ref{sec:survey}.

\section{TwinArch}
\label{sec:twinarch}
TwinArch incorporates the architectural elements identified through the literature review, further refined based on insights from project researchers and mapped onto the three selected DT platforms. Designed for practitioners and researchers involved in the design and development of DTs, TwinArch offers scientifically sound and practical UML artifacts that can be customized to support the instantiation and implementation of new DTs in specific domains or serve as a guideline for documenting existing DTs.

The remainder of this Section is organized as follows. Section \ref{sec:archstructure} provides an overview of TwinArch's structure. Sections \ref{sec:modview} and \ref{sec:compview} delve into the module and component views of TwinArch, respectively. Section \ref{sec:traceview} outlines the traceability view, linking the architectural elements of the aforementioned views. Lastly, Section \ref{sec:dynview} presents the dynamic view of two use cases, i.e., state monitoring and prediction. The complete architecture documentation is accessible on the TwinArch website: \url{https://alessandrasomma28.github.io/twinarch/}.

 \subsection{Structure Overview}
\label{sec:archstructure}
Figure \ref{fig:twinarchstructure} presents an overview of the TwinArch's structure. In line with the SEI Views and Beyond, the proposed reference architecture is organized into multiple views, each addressing specific aspects of the DT system. 
The \textbf{Module Twin View} (MTV) models the domain entities of DTs using the \textit{UML Class Diagram} notation. It employs decomposition, generalization, and usage styles to define high-level relationships and dependencies among domain classes. At a more detailed level, the \textbf{Component Twin View} (CTV) focuses on the specific components of DTs and their interactions, represented through a \textit{UML Component Diagram}. This view adopts tier-based and shared-data architectural styles to describe the relationships among components, providing a finer-grained representation than the MTV.

The \textbf{Traceability Twin View} (TTV) establishes a mapping between the structural elements of the MTV and CTV using a \textit{Matrix Diagram}. This ensures a traceability path from the high-level domain entities defined in the MTV to the detailed components described in the CTV. Lastly, the \textbf{Dynamic Twin View} (DTV) employs \textit{UML Sequence Diagram} to illustrate the interactions among structural elements (classes or components) at runtime. It provides a dynamic perspective, capturing interactions necessary to fulfill DT functionalities in two distinct use cases, i.e., physical system state monitoring and prediction, which represent two of the most typical key objectives of DTs across various industries.

\subsection{Module Twin View}
\label{sec:modview}
The \textit{Module Twin View} defines the structure of a DT system by organizing it into modules (i.e., entities) and relationships between them. It can be expressed as:

\noindent
\begin{align}
    MTV = \{DTE, ER\}
\end{align}

\begin{table*}[!h]
\centering
\caption{Architectural elements of the Module Twin View: Digital Twin Domain Entities catalog.}
\label{tab:dtde}
\resizebox{\textwidth}{!}{%
\begin{tabular}{|l|p{2.7cm}|p{7cm}|p{4cm}|p{3.3cm}|p{3.3cm}|p{3.3cm}|}
\hline
\textbf{ID} & \textbf{Name} & \textbf{Description} & \textbf{Literature Ref.} & \textbf{Azure Digital Twins} & \textbf{Eclipse Ditto} & \textbf{FIWARE} \\ \hline
$dte_1$ & PhysicalTwin & A real entity to be digitally replicated. & \cite{M01_BlockchainDT24, M03_DTTurbine24, M04_DTDLT24, M05_DTPort22, M06_DTPDM24, M11_DTAnomalyTII23, M13_DTHealth23, M14_AIassitedDT23, M16_ArchitecturePdm23, M17_DTCNC23, M32_ContextAwareDT21, M38_DTSixLayer20, M39_DesignFramework20} & N/A & N/A & N/A \\ \hline

$dte_2$ & DataProvider & A facilitator of data, responsible for transmitting raw data from the physical system to the DT. & 
\cite{M03_DTTurbine24, M04_DTDLT24, M05_DTPort22, M06_DTPDM24, M07_DTWaterPlatform24, M08_SmartCityDT24, M10_SmartSpacesDT24, M11_DTAnomalyTII23, M12_OpenTwins23, M13_DTHealth23, M14_AIassitedDT23, M15_OpenArchitecture23, M16_ArchitecturePdm23, M17_DTCNC23, M22_DTArchitectureFiware22, M25_BuidingGuideFiware22, M27_CognitiveDT22, M28_DTPlatforms22, M32_ContextAwareDT21, M35_ProcessPred21, M37_DTPatternCatalog20, M38_DTSixLayer20, M39_DesignFramework20, M40_ModelingDT21, M41_DTFourLayer, M43_ModelHealthDT24} & N/A & N/A & N/A \\ \hline

$dte_3$ & DataReceiver & A mediator between physical and digital twins, responsible for ensuring the transmission of feedback from the DT to the physical world. & \cite{M10_SmartSpacesDT24, M11_DTAnomalyTII23, M13_DTHealth23, M14_AIassitedDT23, M22_DTArchitectureFiware22, M25_BuidingGuideFiware22, M27_CognitiveDT22, M32_ContextAwareDT21, M37_DTPatternCatalog20, M38_DTSixLayer20, M41_DTFourLayer} & N/A & N/A & N/A \\ \hline

$dte_4$ & Adapter & An information converter, responsible for ensuring compatibility and integration between multiple data sources and the DT. & \cite{M03_DTTurbine24, M08_SmartCityDT24, M12_OpenTwins23, M14_AIassitedDT23, M16_ArchitecturePdm23, M22_DTArchitectureFiware22, M25_BuidingGuideFiware22, M30_KeyComponentsDT22, M34_DTModeling21, M35_ProcessPred21, M37_DTPatternCatalog20, M38_DTSixLayer20, M43_ModelHealthDT24} & Event Routing (\totally) & Protocol Adapter (\totally) & IoT Agent (\totally) \\ \hline

$dte_{5}$ & P2DAdapter & An adapter for physical data, responsible for converting and preparing data for integration into the DT system. & \cite{M03_DTTurbine24, M07_DTWaterPlatform24, M08_SmartCityDT24, M22_DTArchitectureFiware22, M25_BuidingGuideFiware22, M37_DTPatternCatalog20, M43_ModelHealthDT24} & Event Route (\totally) & Connectivity API (\totally) & IoT Agent (\totally) \\ \hline

$dte_{6}$ & D2PAdapter & An adapter for DT data, responsible for converting and preparing data for integration into $dte_1$. & \cite{M08_SmartCityDT24, M22_DTArchitectureFiware22, M25_BuidingGuideFiware22, M37_DTPatternCatalog20, M43_ModelHealthDT24} & Event Grid (\totally) & Connectivity API (\totally) & IoT Agent (\totally) \\ \hline

$dte_7$ & DigitalRepresentation & A digital representation of a real-world entity, responsible for abstracting its key structural and behavioral aspects. & \cite{M01_BlockchainDT24, M02_KnowledgeDT24, M03_DTTurbine24, M07_DTWaterPlatform24, M08_SmartCityDT24, M09_maketwin24, M10_SmartSpacesDT24, M11_DTAnomalyTII23, M13_DTHealth23, M15_OpenArchitecture23,M16_ArchitecturePdm23, M19_ProductLineDT23, M21_AnalysisFramework22, M24_DTDataspace22, M28_DTPlatforms22, M30_KeyComponentsDT22, M31_AutonomicDTs22, M34_DTModeling21, M38_DTSixLayer20, M39_DesignFramework20, M40_ModelingDT21, M41_DTFourLayer, M43_ModelHealthDT24} & \partially  & \partially & \partially \\ \hline

$dte_8$ & DigitalShadow & A collection of temporal data traces, responsible for representing $dte_1$ states grouped by shadow types. & \cite{M03_DTTurbine24, M09_maketwin24, M22_DTArchitectureFiware22, M25_BuidingGuideFiware22, M29_ConceptualizingDT22, M35_ProcessPred21} & Digital Twin Model  (\totally) & Things (\totally) & Context Entities (\totally) \\ \hline

$dte_9$ & ShadowManager & A creator and manager of multiple $dte_8$, responsible for lifecycle management of digital shadows. & \cite{M03_DTTurbine24, M09_maketwin24, M13_DTHealth23, M22_DTArchitectureFiware22, M25_BuidingGuideFiware22, M29_ConceptualizingDT22, M35_ProcessPred21} & Model Management (\totally) & Thing Management (\totally) & Context Broker (\totally) \\ \hline

$dte_{10}$ & DigitalModel & A digital representation of $dte_1$ behavioral aspects, for enabling dynamic simulation. & \cite{M03_DTTurbine24, M04_DTDLT24, M05_DTPort22, M09_maketwin24, M11_DTAnomalyTII23, M13_DTHealth23, M12_OpenTwins23, M15_OpenArchitecture23, M16_ArchitecturePdm23, M19_ProductLineDT23, M21_AnalysisFramework22, M24_DTDataspace22, M28_DTPlatforms22, M29_ConceptualizingDT22, M41_DTFourLayer, M43_ModelHealthDT24} & \notatall & \notatall  & \notatall \\ \hline

$dte_{11}$ & ModelManager & A creator and manager of multiple $dte_{10}$, responsible for integrating and synchronizing multiple digital models. & \cite{M03_DTTurbine24, M04_DTDLT24, M09_maketwin24, M13_DTHealth23, M29_ConceptualizingDT22, M43_ModelHealthDT24} & \notatall & \notatall & \notatall \\ \hline

$dte_{12}$ & TwinManager & A central orchestrator to $dte_{9}$ and $dte_{11}$ combined functionalities, responsible for cohesive management. & \cite{M03_DTTurbine24, M04_DTDLT24, M05_DTPort22, M06_DTPDM24, M09_maketwin24, M13_DTHealth23, M29_ConceptualizingDT22, M31_AutonomicDTs22, M41_DTFourLayer} & Model Management (\partially) & Thing Management (\partially)  & Context Broker (\partially) \\ \hline

$dte_{13}$ & ServiceManager & A creator of DT services, responsible for managing and executing DT services. & \cite{M03_DTTurbine24, M04_DTDLT24, M05_DTPort22, M06_DTPDM24, M07_DTWaterPlatform24, M08_SmartCityDT24, M09_maketwin24, M11_DTAnomalyTII23, M16_ArchitecturePdm23, M17_DTCNC23, M18_MDDT23, M21_AnalysisFramework22, M22_DTArchitectureFiware22, M24_DTDataspace22, M31_AutonomicDTs22, M38_DTSixLayer20, M41_DTFourLayer} & Azure Stream Analytics (\totally)  & Event Handling (\partially) & Perseo (\partially) \\ \hline

$dte_{14}$ & FeedbackProvider & A generator of alerts, events, and commands, responsible for channeling feedback from the DT to the $dte_1$. & \cite{M04_DTDLT24, M10_SmartSpacesDT24, M11_DTAnomalyTII23, M13_DTHealth23, M14_AIassitedDT23, M27_CognitiveDT22, M29_ConceptualizingDT22, M30_KeyComponentsDT22, M32_ContextAwareDT21, M34_DTModeling21} & \notatall & \notatall & \notatall \\ \hline

$dte_{15}$ & DataManager & An aggregator of data circulating within the DT, responsible for efficient management, storage, and retrieval. & \cite{M03_DTTurbine24, M04_DTDLT24, M06_DTPDM24, M07_DTWaterPlatform24, M08_SmartCityDT24, M09_maketwin24, M11_DTAnomalyTII23, M12_OpenTwins23, M13_DTHealth23, M14_AIassitedDT23, M15_OpenArchitecture23, M16_ArchitecturePdm23, M17_DTCNC23, M21_AnalysisFramework22, M22_DTArchitectureFiware22, M25_BuidingGuideFiware22, M28_DTPlatforms22, M30_KeyComponentsDT22, M31_AutonomicDTs22, M32_ContextAwareDT21, M34_DTModeling21, M45_CloudDT23} & Azure Event Hub (\totally) & Thing Management (\partially) & Context Broker and QuantumLeap (\totally)  \\ \hline

$dte_{16}$ & DataModel & A model representing the logical structure of exchanged data, for ensuring data interoperability. & \cite{M01_BlockchainDT24, M02_KnowledgeDT24, M08_SmartCityDT24, M12_OpenTwins23, M21_AnalysisFramework22, M22_DTArchitectureFiware22, M24_DTDataspace22, M25_BuidingGuideFiware22, M27_CognitiveDT22} & Interoperability DTDL Models (\totally) & Thing Management (\totally) & Smart Data Models (\totally) \\
\hline
\end{tabular}%
}
\end{table*}
where: 
\begin{itemize}[leftmargin=5mm, topsep=0.5pt]
    \item \textit{DTE} represents the set of \textbf{Digital Twin Domain Entities}, which encapsulate the structural and functional characteristics of specific domain elements, such as physical twin or digital models, forming the foundational elements for building a DT system. The final set of DTEs, identified through the systematic review and refined through the specific search conducted on available DT platforms and solutions, is summarized in Table \ref{tab:dtde}.
    
    The table provides DTE's IDs, names, descriptions and the literature references from which they have been derived. Moreover, it reports the mapping of each DTE onto the corresponding software elements belonging to the DT platforms selected as reference for this study, namely FIWARE, Eclipse Ditto and Azure Digital Twins. In the table, in particular, mappings are represented using three symbols: \totally, \partially, and \notatall. A \ding{51} symbol indicates that the DTE class can be fully implemented using the tools offered by the respective DT platform. A \(\sim\) symbol denotes that the DTE class is only partially implementable with the existing tools, necessitating additional resources to complete the element. Lastly, a \ding{55} symbol signifies that the DTE class responsibilities cannot be addressed using the tools provided by the DT platform.
    
    \item $ER$ denotes the set of \textbf{Entity Relationships} that describe connections among DT domain entities. 
    These relationships are established based on the architectural styles adopted in the MTV design. 
    \begin{itemize}[label=\textbf{$\diamond$},leftmargin=5mm, topsep=0pt]
        \item The \textit{Decomposition Style} uses a divide-and-conquer approach to manage the system's complexity by breaking it into smaller modules, introducing the \texttt{is-part-of relationship}. This relationship can either represent a strong composition, where the part cannot exist independently of the whole, or an aggregation, where the part can exist independently of the whole. \item The \textit{Generalization Style} models common functionalities across modules to promote sharing and reuse, introducing the \texttt{is-a relationship}. 
        \item The \textit{Uses Style} models dependencies between modules, supporting incremental design and introducing the \texttt{use relationship}.  
        \item The \texttt{abstraction} relationship is introduced to represent the connection between the domain entity representing the physical system and its corresponding virtual counterpart.
    \end{itemize}
\end{itemize}

\begin{figure*}[t!]
\centerline{\includegraphics[width=0.7\textwidth]{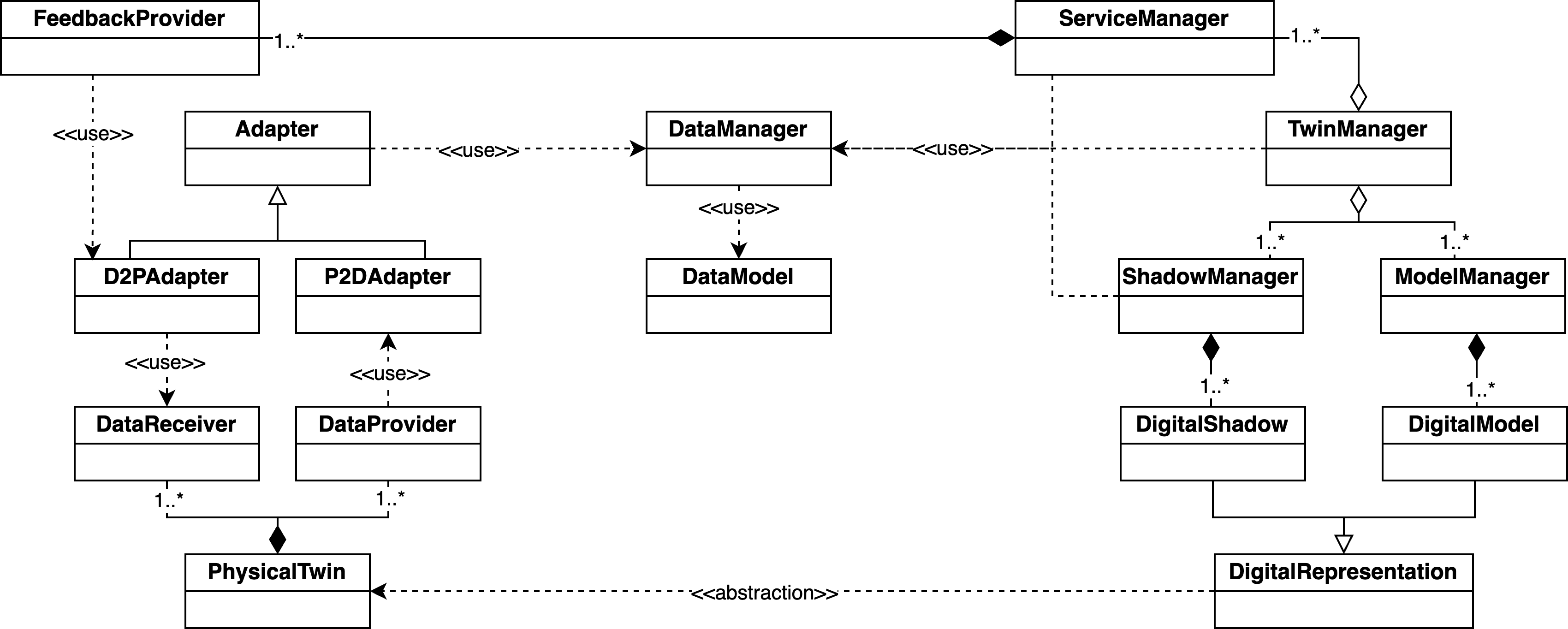}}
\caption{Module Twin View: UML Class Diagram.}
\label{fig:classdiagram}
\end{figure*}

Figure \ref{fig:classdiagram} depicts the \textit{UML Class Diagram} illustrating the Digital Twin Domain Entities as classes, along with their interrelationships. \texttt{Use} and \texttt{abstraction} relationships are depicted in the diagram through arrows to which the corresponding stereotypes are applied. As for the \texttt{is-part-of} relationship, composition is represented with a filled diamond at the end of the association line that connects to the whole, while aggregation is depicted with an empty diamond. Relationship multiplicity specifies how many instances of a part can be associated with a single instance of the whole. Finally, the generalization relationship is represented as a solid line with a hollow triangle arrowhead pointing towards the more general (parent) class. 

The following paragraphs illustrate the architectural elements of the Module Twin View providing a description of their main functions and on their mutual relationships. Moreover, in yellow boxes the reader can find a discussion on whether and how each entity is mapped to the considered software technologies. When applicable, relevant examples of the mapping are also discussed.

\vspace{3em}
\noindent \textbf{PhysicalTwin and DigitalRepresentation.}

\noindent \texttt{PhysicalTwin} represents the real-world entity that is digitally replicated within the DT system. It acts as source of truth, providing essential data and state information to its virtual counterpart for various use cases, including simulation, monitoring, and prediction. 

\texttt{DigitalRepresentation}, on the other hand, abstracts the structural and behavioral characteristics of \texttt{PhysicalTwin}, accommodating different levels of granularity. The \texttt{abstraction} relationship ensures that  \texttt{DigitalRepresentation} captures the key properties and functionalities that are relevant to the objectives of the DT system. For example, for system monitoring, it may include structural and behavioral features like dimensions and states, while omitting irrelevant attributes such as aesthetic details. Additionally, abstraction allows \texttt{DigitalRepresentation} to incorporate derived or aggregated data, such as maintenance history or performance metrics, enhancing its value and utility.

As illustrated in Fig. \ref{fig:classdiagram} and discussed in detail later, \texttt{DigitalRepresentation} can be further classified into digital shadows and digital models. 

\begin{boxC}
\footnotesize
\texttt{PhysicalTwin} exists solely in the physical domain and is not mapped to any platform, as the discussed technologies focus on digital aspects. In contrast, \texttt{DigitalRepresentation} is only partially supported, with the selected technologies addressing specific aspects of its specialized forms, such as shadowing entities, rather than fully covering the entire \texttt{DigitalRepresentation}.
\end{boxC}


\noindent \textbf{DataProvider, DataReceiver and Adapters}. 

\noindent  \texttt{DataProvider} acts as an intermediary between the physical and digital twins, generating the flow of information from the physical world into the DT system. It ensures that data from the \texttt{PhysicalTwin} are effectively transmitted to the digital counterpart. 
\texttt{DataReceiver} operates in the opposite direction, mediating the flow of information and feedback from the DT to the physical world. Together, \texttt{DataProvider} and \texttt{DataReceiver} maintain bidirectional synchronization between the physical and digital spaces. 
As depicted, \texttt{DataProvider} and \texttt{DataReceiver} are modelled as \texttt{a-part-of} \texttt{PhysicalTwin} by a strong composition relationship, meaning that the lifetimes of \texttt{DataProvider} and \texttt{DataReceiver} are encompassed within the lifetime of  \texttt{PhysicalTwin}. 

\texttt{Adapter} performs the transformations necessary for seamless data exchange between multiple and heterogeneous data sources and the DT system. \texttt{Adapter} class is specialized into \texttt{P2DAdapter} and \texttt{D2PAdapter}, which manage data flows in specific directions. \texttt{P2DAdapter} focuses on physical-to-digital data flows, transforming and preparing data sent by  \texttt{DataProvider} for integration into the DT system. Conversely, \texttt{D2PAdapter} handles digital-to-physical data flows, adapting and preparing feedback or commands provided by the DT system for \texttt{DataReceiver}.

As shown in Fig. \ref{fig:classdiagram}, there is a usage relationship from \texttt{DataProvider} to \texttt{P2DAdapter} indicating that the former relies on the functions offered by the latter: in fact, data retrieved by \texttt{DataProvider} from \texttt{PhysicalTwin} are processed by \texttt{P2DAdapter} and transformed into a format compatible with the DT system, defined by \texttt{DataModel} class, prior to be transmitted to \texttt{DataManager}. A similar usage relationship holds from \texttt{D2PAdapter} to \texttt{DataReceiver} that involves an opposite data flow, omitted for brevity. 

\begin{boxC}
\footnotesize
\texttt{DataProvider} and \texttt{DataReceiver} are not directly mapped to any of the selected platforms, as they represent physical-world sensors and actuators. Instead, \texttt{P2DAdapter} and \texttt{D2PAdapter} are supported by Event Routing tools in Azure Digital Twins, the Connectivity API in Eclipse Ditto, and the IoT Agents available in FIWARE.

\vspace{1em}
\textbf{Exemplars in DT Platforms}: 
Let us consider a traffic loop sensor that measures vehicle flow and periodically transmits its readings to the Digital Twin system for integration into the digital model. Below, we examine how this sensor interacts with adapters within the Azure Digital Twins, Eclipse Ditto, and FIWARE platforms.

\vspace{1em}
In \textit{\textbf{Eclipse Ditto}}, \texttt{P2DAdapter} leverages the \textit{Connectivity API}\footnote{\url{https://eclipse.dev/ditto/basic-connections.html}} to handle data streams from the traffic loop sensor. An example request to the Ditto Connectivity API is described below:
\begin{lstlisting}[style=bashstyle]
    curl -X POST 
    'http://ditto/connectivity?filter=type=trafficLoop' 
    -d '{"input": "vehicleCount", 
        "output": "processedFlow"}'
\end{lstlisting}
The filter ensures that only data from traffic loop sensors are processed. The payload specifies the transformation from raw vehicle count data to a processed flow representation. The adapter retrieves the traffic flow data, enriches them with additional metadata (e.g., timestamp, location), and integrates them into the Ditto-managed Twin in a structured format.

\vspace{1em}
In\textit{ \textbf{Azure Digital Twin}}, \texttt{P2DAdapter} is implemented using \textit{Event Routes}\footnote{\url{https://learn.microsoft.com/en-us/azure/digital-twins/concepts-route-events}}, which route telemetry data from the sensor to the target Twin. A sample configuration for an Event Route is as follows:
\begin{lstlisting}[style=jsonstyle]
    { "id": "trafficToTwinRoute",
      "source": "/eventhub/telemetry",
      "target": "/digitalTwins/trafficTwin",
      "filter": "$event.properties.sensorType == 'trafficLoop'" }
    \end{lstlisting}
This configuration defines the route’s unique identifier, the telemetry data source, and the target Twin. A filter ensures only traffic loop data are processed. The traffic loop sensor sends vehicle flow data via Azure Event Hubs\footnote{\url{https://azure.microsoft.com/en-us/products/event-hubs}}. The Event Route processes the readings and transforms them into a JSON format compatible with the schema of the target Twin.

\vspace{1em}
In \textit{\textbf{FIWARE}}, \texttt{P2DAdapter} is implemented using an IoT Agent, such as \textit{IoT Agent Ultralight 2.0}\footnote{\url{https://github.com/telefonicaid/iotagent-ul}}, which converts lightweight payloads into NGSI-LD context updates. A sample interaction is the following:
\begin{lstlisting}[style=bashstyle]
curl -iX POST 
'http://<iot-agent-host>:<port>/iot/d?k=<apikey>
        &i=<device-id>' 
-H 'Content-Type: text/plain' -d 'f|35'
\end{lstlisting}
In this case, the IoT Agent firstly authenticates the device using the API key and the identifier and then 
translates the received payload (\texttt{f|35}, where \texttt{f} denotes traffic flow and \texttt{35} the number of vehicles measured) into an NGSI-LD compliant structure. This transformed context is forwarded to the FIWARE Context Broker, enabling it to store and process the sensor data within the DT system.
\end{boxC}

\noindent \textbf{DigitalShadow and ShadowManager.} 

\noindent \texttt{DigitalShadow} is a specialization of \texttt{DigitalRepresentation}, focusing on data-related aspects of  \texttt{PhysicalTwin}. 

It represents a collection of temporal data traces that capture the states of \texttt{PhysicalTwin} over time. This data-centric representation is grouped by shadow types, enabling advanced functionalities such as anomaly detection, predictive maintenance, and historical analysis. 

\texttt{ShadowManager} oversees the lifecycle of multiple \texttt{Digital Shadow} instances, including their creation, updates, and deletion. It organizes shadows by temporal properties and types, maintaining their integrity and consistency and ensuring seamless coordination across DT system elements. The relationship between  \texttt{ShadowManager} and \texttt{DigitalShadow} is a strong composition (is-part-of) relationship, as the lifetime of the shadow instances is dependent on the lifetime of the Manager.

\begin{boxC}
\footnotesize
\texttt{DigitalShadow} and \texttt{ShadowManager} are both supported by the selected DT platforms. In Azure Digital Twins, \texttt{DigitalShadow} is implemented using Digital Twin Model to store data traces, while model management tools support  \texttt{ShadowManager} for lifecycle handling. Eclipse Ditto models \texttt{DigitalShadow} as Things and uses Thing Management for managing multiple shadows. FIWARE represents shadows using NGSI-LD Context Entities and employs the Context Broker to manage their lifecycle.
\end{boxC}

\noindent \textbf{DigitalModel and ModelManager.} 

\noindent \texttt{DigitalModel} is the digital representation of the behavioral aspects of \texttt{PhysicalTwin}. It focuses on modeling the operational behavior of the physical entity, enabling dynamic simulations and predictive analysis. While \texttt{DigitalShadow} captures the historical states of \texttt{PhysicalTwin}, \texttt{DigitalModel} complements this by simulating current and future states, offering insights into system performance under different conditions, and enabling scenario-based analysis. 

\texttt{ModelManager} is responsible for overseeing and managing multiple \texttt{DigitalModel} instances. It facilitates the integration, synchronization, and orchestration of these models, ensuring cohesive and accurate behavioral simulations. Additionally,  \texttt{ModelManager} ensures the consistency of the models, aligning them with the corresponding \texttt{PhysicalTwin} and maintaining their integrity within the broader DT system. For this reason, the relationship between  \texttt{ModelManager} and  \texttt{DigitalModel} is a strong composition (is-part-of) relationship.

\begin{boxC}
\footnotesize
\texttt{DigitalModel} and \texttt{ModelManager} are not natively supported by the selected platforms, as these primarily address the structural modeling of physical assets and lack features for behavioral modeling. Implementing these entities requires the integration of specialized simulation tools. For instance, MATLAB Simulink can model and simulate dynamic systems, while Eclipse SUMO (Simulation of Urban Mobility) is suitable for traffic and urban planning simulations. Other examples include AnyLogic for multi-method modeling and Python-based frameworks such as OpenModelica for system-level simulations.
\end{boxC}

\noindent \textbf{TwinManager.} 

\noindent \texttt{TwinManager} serves as the central orchestrator, managing and coordinating the functionalities of both \texttt{ShadowManager} and \texttt{ModelManager}. It ensures the cohesive operation of the DT system by aligning the data-driven aspects represented by digital shadows with the simulation-driven aspects captured by digital models. Additionally, \texttt{TwinManager} handles cross-functional tasks such as synchronizing data between shadows and models. 

\begin{boxC}
\footnotesize
\texttt{TwinManager} is partially supported by the selected platforms because it manages both data-driven flows from  \texttt{ShadowManager}, which are fully supported, and simulation-driven flows from \texttt{ModelManager}, which are not supported. Azure Digital Twins provides Twin Management for orchestrating operations, Eclipse Ditto offers Ditto Management for device interactions and shadow functionalities, and FIWARE uses the Context Broker for synchronizing and integrating services.
\end{boxC}

\noindent \textbf{ServiceManager and FeedbackProvider.} 

\noindent \texttt{ServiceManager} is responsible for implementing, coordinating, and managing the services offered by the DT system, such as monitoring, anomaly detection, and prediction. \texttt{FeedbackProvider} produces the DT feedback, including alerts, events, and commands, directly linked to the services managed by \texttt{ServiceManager}.  
The operations of both entities enable the DT to react to observed conditions, issue alerts, and send commands to influence the behavior of \texttt{PhysicalTwin}. As critical entities, they support closed-loop operations, providing real-time feedback to drive adaptive responses in the physical domain. The relationship between  \texttt{ServiceManager} and \texttt{FeedbackProvider} is a strong composition (is-part-of) relationship.

\begin{boxC}
\footnotesize
\texttt{ServiceManager} is fully supported by Azure DTs, leveraging Azure Stream Analytics to handle service-related tasks effectively. In comparison, the other two platforms offer partial service management capabilities.  Eclipse Ditto facilitates service orchestration through its management and integration features, including event handling, while FIWARE supports real-time alerts and actions using tools such as Perseo. \texttt{FeedbackProvider} is not natively supported on any of the selected platforms, necessitating custom implementation to develop specialized feedback generation mechanisms tailored to the implemented services.
\end{boxC}

\noindent \textbf{DataManager and DataModel.} 

\noindent \texttt{DataManager} is responsible for aggregating, managing, and retrieving data within the DT system, ensuring efficient storage, consistency, and integration. It relies on \texttt{DataModel}, which defines the logical structure of the data, facilitating interoperability by standardizing data formats for internal and external exchanges. 
 
\texttt{DataManager} is a pivotal entity utilized by various DTEs. For instance, \texttt{TwinManager} uses it to manage data flows between shadows and models, \texttt{ServiceManager} depends on it for service-related data operations, and \texttt{ShadowManager} leverages it to store temporal data traces representing \texttt{PhysicalTwin}’ states.

\begin{boxC}
\footnotesize
\texttt{DataManager} and \texttt{DataModel} are fully supported by the Azure Digital Twins and FIWARE platforms. Azure DTs utilizes the Digital Twin Definition Language for data modeling and Azure Event Hub to address critical data management functions. FIWARE employs Smart Data Models for data definitions and combines the Context Broker with QuantumLeap to facilitate efficient data storage and management. In contrast, Eclipse Ditto offers comprehensive support for data modeling through its Things Management API, which provides only partial support for data management functionalities.

\vspace{1em}
\textbf{Exemplars in DT Platforms}: Let us consider the same traffic loop sensor measuring vehicle flow of the previous example. The exchanged measurement are modeled by the selected platforms as explained below.

\vspace{1em}
In \textit{\textbf{Eclipse Ditto}}, the \textit{Things Management} API\footnote{\url{https://eclipse.dev/ditto/basic-thing.html}} is used to define and manage digital representations of physical devices (the Things). 
For example, a traffic loop sensor can be represented as a Thing with a unique identifier, thingId, and an attribute, vehicleCount, which stores the observed traffic flow as an integer value. The following JSON schema demonstrates how the sensor can be modeled:
\begin{lstlisting}[style=jsonstyle]
{ "thingId": "example:TrafficSensor",
  "attributes": {
    "vehicleCount": {
      "type": "integer",
      "value": 35
    } } }
\end{lstlisting}

\vspace{1em}
In \textit{\textbf{Azure Digital Twins}}, the \textit{DTDL}\footnote{\url{https://azure.github.io/opendigitaltwins-dtdl/DTDL/v3/DTDL.v3.html}} is adopted to define the structure of Digital Twin models. This language allows for creating detailed representations of physical entities, including their properties, telemetry, commands, and relationships. For instance, a traffic loop sensor can be modeled as an interface in DTDL, representing its data collection functionality, as illustrated below:
\begin{lstlisting}[style=jsonstyle]
{ "@id": "dtmi:example:TrafficSensor;1",
  "@type": "Interface",
  "contents": [ {
      "@type": "Telemetry",
      "name": "vehicleCount",
      "schema": "integer"
    } ] }
\end{lstlisting}
Data collected by Azure Event Hub can be seamlessly stored and analyzed using Azure services such as Azure Data Lake or Cosmos DB, facilitating advanced data-driven insights and operations.

\vspace{1em}
In \textit{\textbf{FIWARE}}, \textit{Smart Data Models}\footnote{\url{https://github.com/smart-data-models}} are employed for data definition and the \textit{Context Broker} for data management. Traffic loop sensor data can be represented using an NGSI-LD schema based on the Transportation data model. For instance, the sensor could be represented as an entity of type \textit{TrafficFlowObserved}, with a unique identifier and attributes capturing its observations:
\begin{lstlisting}[style=jsonstyle]
{ "id": "urn:ngsi-ld:TrafficFlowObserved:TLF01",
  "type": "TrafficFlowObserved",
  "location": {
    "type": "Point",
    "coordinates": [40.7128, -74.0060]
  },
  "vehicleFlow": {
    "value": 35,
    "observedAt": "2024-12-10T12:00:00Z"}}
\end{lstlisting}
\end{boxC}

\subsection{Component Twin View}
\label{sec:compview}

\begin{table*}[!h]
\centering
\caption{Architectural elements of the Component Twin View: Digital Twin Components catalog.}
\label{tab:dtc}
\resizebox{\textwidth}{!}{%
\begin{tabular}{|l|p{2.5cm}|p{7.3cm}|p{4.4cm}|p{3cm}|p{3cm}|p{3cm}|}
\hline
\textbf{ID} & \textbf{Name} & \textbf{Description} & \textbf{Literature Ref.} & \textbf{Azure Digital Twins} & \textbf{Eclipse Ditto} & \textbf{FIWARE} \\ \hline
$dtc_1$ & PhysicalTwin & A real-world asset to be replicated by the Digital Twin. & \cite{M01_BlockchainDT24, M03_DTTurbine24, M04_DTDLT24, M05_DTPort22, M06_DTPDM24, M10_SmartSpacesDT24, M11_DTAnomalyTII23, M13_DTHealth23, M18_MDDT23, M20_ArchitectingDTDD23, M27_CognitiveDT22, M29_ConceptualizingDT22, M32_ContextAwareDT21, M33_SelfAdaptiveDT21, M35_ProcessPred21, M42_MDDT20, M44_DTHealthEng23} & N/A & N/A & N/A \\ \hline
$dtc_2$ & DataProvider & An intermediary facilitating the transmission of raw data from the physical to the Digital Twin.  & \cite{M01_BlockchainDT24, M03_DTTurbine24, M04_DTDLT24, M05_DTPort22, M06_DTPDM24, M10_SmartSpacesDT24, M11_DTAnomalyTII23, M13_DTHealth23, M14_AIassitedDT23, M18_MDDT23, M20_ArchitectingDTDD23, M27_CognitiveDT22, M29_ConceptualizingDT22, M32_ContextAwareDT21, M33_SelfAdaptiveDT21, M35_ProcessPred21, M42_MDDT20, M44_DTHealthEng23} & N/A & N/A & N/A \\ \hline
$dtc_3$ & DataReceiver & A receiver ensuring feedback, updates, or commands from the $dtc_{22}$ reach the Physical Twin. & \cite{M08_SmartCityDT24, M22_DTArchitectureFiware22, M25_BuidingGuideFiware22, M37_DTPatternCatalog20, M43_ModelHealthDT24} & N/A & N/A & N/A \\ \hline
$dtc_4$ & P2DAdapter & A converter that translates physical system data into formats usable by the DT. & \cite{M03_DTTurbine24, M07_DTWaterPlatform24, M08_SmartCityDT24, M22_DTArchitectureFiware22, M25_BuidingGuideFiware22, M37_DTPatternCatalog20, M43_ModelHealthDT24} & Event Route (\totally) & Connectivity API (\totally) & IoT Agent (\totally) \\ \hline
$dtc_5$ & D2PAdapter & A translator that transforms Digital Twin outputs into formats usable by $dtc_1$. & \cite{M08_SmartCityDT24, M22_DTArchitectureFiware22, M25_BuidingGuideFiware22, M37_DTPatternCatalog20, M43_ModelHealthDT24} & Event Grid (\totally) & Connectivity API (\totally) & IoT Agent (\totally) \\ \hline
$dtc_6$ & DataProcessor & A processing unit that filters and organizes raw data, preparing them for integration into the DT system. &\cite{M01_BlockchainDT24, M02_KnowledgeDT24, M03_DTTurbine24, M04_DTDLT24, M05_DTPort22, M06_DTPDM24, M09_maketwin24, M10_SmartSpacesDT24, M11_DTAnomalyTII23, M12_OpenTwins23, M13_DTHealth23, M15_OpenArchitecture23, M18_MDDT23, M19_ProductLineDT23, M21_AnalysisFramework22, M26_IoTwins22, M27_CognitiveDT22, M32_ContextAwareDT21, M33_SelfAdaptiveDT21, M35_ProcessPred21, M42_MDDT20, M44_DTHealthEng23} & Azure Event Hub (\totally) & Thing Management (\totally) & Context Broker (\totally) \\ \hline
$dtc_7$ & StorageManager &  A component that organizes and manages shared data repository $dtc_9$ for efficient storage and retrieval. & \cite{M04_DTDLT24, M07_DTWaterPlatform24, M08_SmartCityDT24, M09_maketwin24, M11_DTAnomalyTII23, M12_OpenTwins23, M13_DTHealth23, M22_DTArchitectureFiware22, M25_BuidingGuideFiware22, M26_IoTwins22, M31_AutonomicDTs22, M32_ContextAwareDT21, M33_SelfAdaptiveDT21, M34_DTModeling21, M35_ProcessPred21, M42_MDDT20} & Azure Event Hub (\totally) & Thing Management (\partially) & Context Broker (\totally) \\ \hline
$dtc_8$ & DataManager & A centralized component ensuring data consistency and availability, aggregating $dtc_6$ and $dtc_7$ functionalities. & \cite{M01_BlockchainDT24, M02_KnowledgeDT24, M04_DTDLT24, M11_DTAnomalyTII23, M13_DTHealth23, M14_AIassitedDT23, M16_ArchitecturePdm23, M22_DTArchitectureFiware22, M25_BuidingGuideFiware22, M26_IoTwins22, M27_CognitiveDT22, M33_SelfAdaptiveDT21, M35_ProcessPred21, M42_MDDT20} & Azure Event Hub (\totally) & Thing Management (\partially) & QuantumLeap (\partially) \\ \hline
$dtc_9$ & SharedStorage & A data accumulator, responsible for storing heterogeneous data from both the physical and digital twins. & \cite{M01_BlockchainDT24, M02_KnowledgeDT24, M03_DTTurbine24, M04_DTDLT24, M07_DTWaterPlatform24, M08_SmartCityDT24, M09_maketwin24, M11_DTAnomalyTII23, M12_OpenTwins23, M13_DTHealth23, M14_AIassitedDT23, M15_OpenArchitecture23, M16_ArchitecturePdm23, M18_MDDT23, M22_DTArchitectureFiware22, M25_BuidingGuideFiware22, M26_IoTwins22, M27_CognitiveDT22, M31_AutonomicDTs22, M32_ContextAwareDT21, M33_SelfAdaptiveDT21, M34_DTModeling21, M35_ProcessPred21, M42_MDDT20, M44_DTHealthEng23, M45_CloudDT23} & Azure Data Lake or Cosmos DB (\totally) & MongoDB (\partially) & MongoDB, TimescaleDB, CrateDB (\partially) \\ \hline 
$dtc_{10}$ & ShadowManager & A component responsible for creating, managing, and overseeing the lifecycle of multiple digital shadows.& \cite{M04_DTDLT24, M09_maketwin24, M12_OpenTwins23, M22_DTArchitectureFiware22, M25_BuidingGuideFiware22, M26_IoTwins22, M27_CognitiveDT22, M31_AutonomicDTs22, M18_MDDT23, M21_AnalysisFramework22, M29_ConceptualizingDT22, M33_SelfAdaptiveDT21, M35_ProcessPred21} & Model Management (\totally) & Ditto Management (\totally) & Context Broker (\totally) \\ \hline
$dtc_{11}$ & ModelManager & A component responsible for creating and managing multiple digital models to simulate different aspects. & \cite{M03_DTTurbine24, M09_maketwin24, M12_OpenTwins23, M23_DTFramework22, M29_ConceptualizingDT22} & \notatall & \notatall & \notatall \\ \hline
$dtc_{12}$ & ModelEngine & A processing unit of digital models, responsible for executing simulations and generating results based on the modeled scenarios. & \cite{M03_DTTurbine24, M09_maketwin24, M12_OpenTwins23, M23_DTFramework22, M29_ConceptualizingDT22} & \notatall & \notatall & \notatall \\ \hline
$dtc_{13}$ & Simulator & A virtualizer of real-world systems, responsible for simulating the behavior of the physical system under various conditions. & \cite{M03_DTTurbine24, M04_DTDLT24, M07_DTWaterPlatform24, M09_maketwin24, M11_DTAnomalyTII23, M15_OpenArchitecture23, M16_ArchitecturePdm23, M20_ArchitectingDTDD23, M21_AnalysisFramework22, M27_CognitiveDT22, M28_DTPlatforms22, M30_KeyComponentsDT22, M31_AutonomicDTs22, M38_DTSixLayer20} & \notatall & \notatall & \notatall \\ \hline
$dtc_{14}$ & TwinManager & An orchestrator synchronizing the functionalities of shadow and model managers with the service-related components for cohesive DT operations. & \cite{M12_OpenTwins23, M13_DTHealth23, M23_DTFramework22, M27_CognitiveDT22, M33_SelfAdaptiveDT21, M35_ProcessPred21, M37_DTPatternCatalog20, M36_AgentArchitecture21, M42_MDDT20} & Twin Management (\partially) & Ditto Management (\partially) & Context Broker (\partially) \\ \hline
$dtc_{15}$ & StateMonitor & A monitoring component, responsible for collecting and forwarding real/simulated states to other components for further analysis or action. & \cite{M03_DTTurbine24, M11_DTAnomalyTII23, M12_OpenTwins23, M13_DTHealth23, M18_MDDT23, M19_ProductLineDT23, M20_ArchitectingDTDD23, M26_IoTwins22, M27_CognitiveDT22, M32_ContextAwareDT21, M36_AgentArchitecture21, M37_DTPatternCatalog20} & Azure Stream Analytics (\totally) & Event Handling (\partially) & FIWARE Cosmos (\totally) \\ \hline
$dtc_{16}$ & DeviationDetector & A comparison unit, responsible for identifying deviations by comparing real or predicted states with expected states to detect any deviation. & \cite{M06_DTPDM24, M10_SmartSpacesDT24, M11_DTAnomalyTII23, M16_ArchitecturePdm23, M27_CognitiveDT22, M35_ProcessPred21, M36_AgentArchitecture21, M37_DTPatternCatalog20, M45_CloudDT23}  & Multivariate Detection toolkit (\totally)  & \notatall & FIWARE Perseo (\partially) \\ \hline
$dtc_{17}$ & Predictor &  A forecasting component, responsible for using current and historical data to anticipate potential future states of the physical system. & \cite{M06_DTPDM24, M07_DTWaterPlatform24, M08_SmartCityDT24, M10_SmartSpacesDT24, M12_OpenTwins23, M14_AIassitedDT23, M16_ArchitecturePdm23, M22_DTArchitectureFiware22, M23_DTFramework22, M25_BuidingGuideFiware22, M35_ProcessPred21, M36_AgentArchitecture21} & Azure ML and DL (\totally) & \notatall & FIWARE Cosmos (\partially) \\ \hline
$dtc_{18}$ & Analyzer & A detailed analytical component, responsible for analyzing real, predicted and simulated states to extract meaningful insights. & \cite{M06_DTPDM24, M07_DTWaterPlatform24, M08_SmartCityDT24, M10_SmartSpacesDT24, M11_DTAnomalyTII23, M12_OpenTwins23, M14_AIassitedDT23, M16_ArchitecturePdm23, M18_MDDT23, M19_ProductLineDT23, M21_AnalysisFramework22, M23_DTFramework22, M27_CognitiveDT22, M28_DTPlatforms22, M29_ConceptualizingDT22, M30_KeyComponentsDT22, M31_AutonomicDTs22, M32_ContextAwareDT21, M33_SelfAdaptiveDT21, M35_ProcessPred21, M36_AgentArchitecture21, M42_MDDT20, M44_DTHealthEng23} & Azure Stream Analytics (\totally) & \partially & FIWARE Perseo and Cosmos (\partially) \\ \hline
$dtc_{19}$ & SolutionFinder & A component responsible for finding the best set of actions to return the system to a desired state after deviation detection. & \cite{M10_SmartSpacesDT24, M11_DTAnomalyTII23, M18_MDDT23, M23_DTFramework22, M33_SelfAdaptiveDT21, M35_ProcessPred21} & \notatall & \notatall & \notatall \\ \hline 
$dtc_{20}$ & ScenarioGenerator & A generator designed to create diverse scenarios, facilitating the preparation and execution of multiple simulations. & \cite{M06_DTPDM24, M09_maketwin24, M23_DTFramework22, M36_AgentArchitecture21} & \notatall & \notatall & \notatall \\ \hline
$dtc_{21}$ & Planner & A planning unit, responsible for developing a solution plan to restore the system to a desired state when deviations or anomalies are detected & \cite{M10_SmartSpacesDT24, M11_DTAnomalyTII23, M18_MDDT23, M19_ProductLineDT23, M29_ConceptualizingDT22, M33_SelfAdaptiveDT21} & \notatall & \notatall & \notatall \\ \hline
$dtc_{22}$ & FeedbackExecutor & A generator of alerts or actionable instructions into the physical system $dtc_1$.  & \cite{M10_SmartSpacesDT24, M11_DTAnomalyTII23, M13_DTHealth23, M18_MDDT23, M19_ProductLineDT23, M27_CognitiveDT22, M29_ConceptualizingDT22, M30_KeyComponentsDT22, M33_SelfAdaptiveDT21, M34_DTModeling21, M35_ProcessPred21} & \notatall & \notatall & \notatall \\ \hline
\end{tabular}%
}
\end{table*}

The \textit{Component Twin View} models the architecture of a DT software system by specifying its components and their connectors. This view provides a finer level of granularity than the Module Twin View, as it emphasizes the internal software components and is more closely aligned with software implementation. The Component Twin View is expressed as:
\begin{equation}
    CTV = \{DTC, CR\}
\end{equation}
where:
\begin{itemize}[leftmargin=5mm, topsep=0.5pt]
    \item $DTC$ represents the set of \textbf{Digital Twin Components}, which are self-contained software entities encapsulating specific functionalities within the DT system. Each component is designed to perform a distinct role, such as data processing, state monitoring, or simulation. Similar to the Module Twin View, the final set of DTCs, identified through the literature review and refined with the platforms specific search, is detailed in Table~\ref{tab:dtc}.

    The table details DTCs identifiers, name, description, references from the literature. Moreover, as done in the Module Twin View, Tab. \ref{tab:dtc} reports the mapping onto the software tools of the selected technologies (Azure Digital Twins, Eclipse Ditto, and FIWARE) specifying whether the mapping is total (\totally), partial (\partially), or not possible (\notatall).
    \item $CR$ denotes the set of \textbf{Component Relationships}, which describe the interactions and dependencies among components. These relationships are defined based on the architectural styles used in the design. 
    \begin{itemize}[label=\textbf{$\diamond$},leftmargin=5mm, topsep=0pt]
        \item The \textit{Tier Style} organizes components into functional groups according to their execution structures, modeling their composition through the \texttt{is-part-of relationship}. 
        \item The \textit{Shared-Data Repository Style} defines the \texttt{usage relationship}, illustrating how components access a shared repository to read or write data. 
     \end{itemize}
    Beyond the above mentioned relationships, the Component Twin View incorporates additional relationships commonly adopted in C\&C views \cite{Book_DocumentingSA}, namely:
    \begin{itemize}[label=\textbf{$\diamond$},leftmargin=5mm, topsep=0pt]
        \item the \texttt{assembly} relationship, which connects a component's required interface to the provided interface of another component. 
        \item the \texttt{port attachment} or association relationship, which models how components exchange information or collaborate, typically visualized as connections between their ports. 
        \item the \texttt{interface delegation} relationship, which links a component's internal ports to its external ports in cases where the component includes a sub-architecture.
         \end{itemize}
\end{itemize}

\begin{figure*}[!ht]
\centerline{\includegraphics[width=0.85\textwidth]{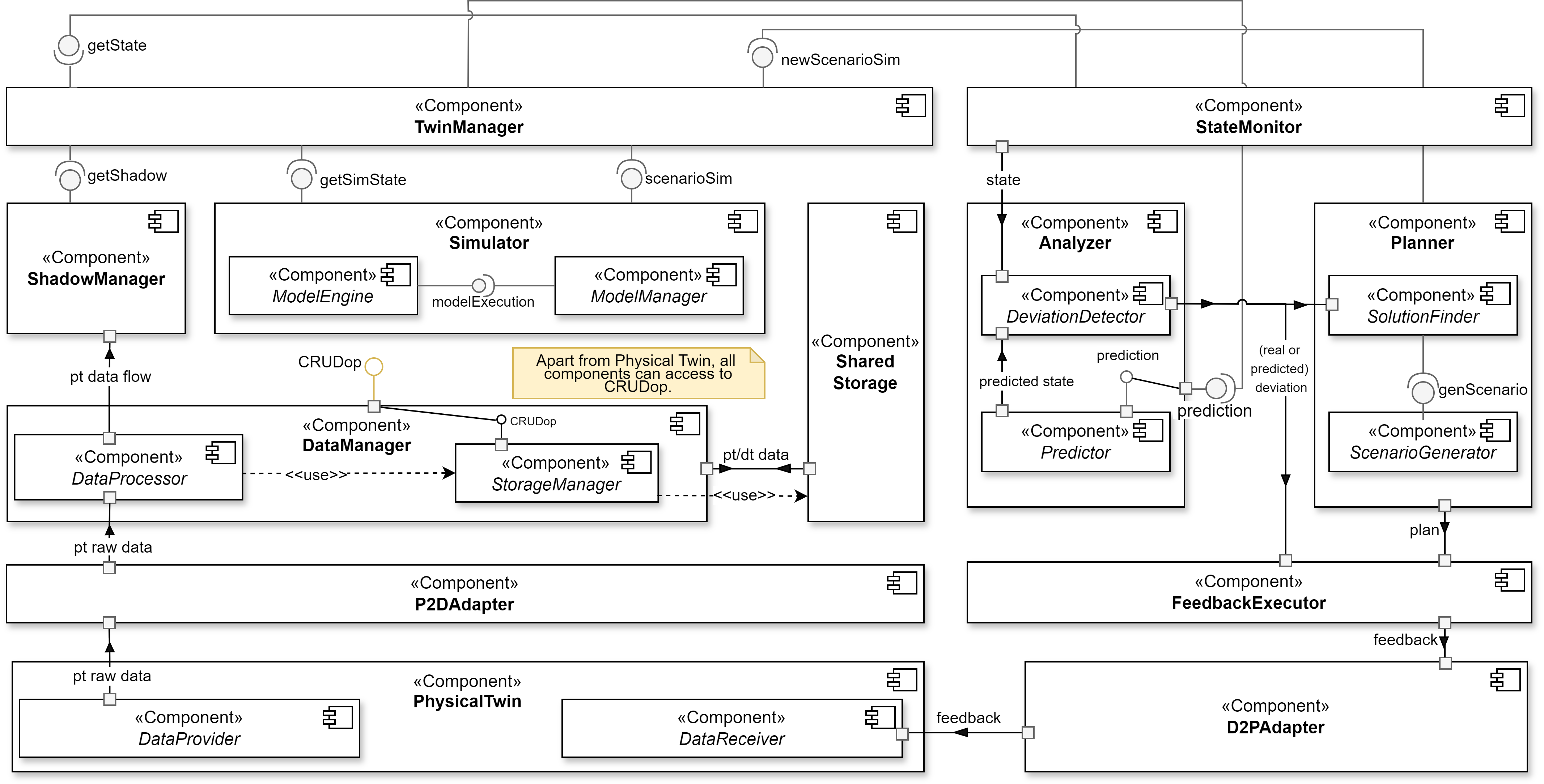}}
\caption{Component Twin View: UML Component Diagram.}
\label{fig:component}
\end{figure*}

Figure \ref{fig:component} presents the \textit{UML Component Diagram,} illustrating the Digital Twin Components and their relationships.  The subsequent paragraphs provide an in-depth explanation of the architectural elements of the Component Twin View, their relationships, and their mappings the entities of the module view. Where applicable, examples of usage of the components in concrete scenarios is also provided, together with specific examples of implementation within the considered platforms.

\vspace{1em}
\noindent \textbf{PhysicalTwin, DataProvider and DataReceiver.} 

\noindent \texttt{PhysicalTwin}, \texttt{DataProvider} and \texttt{DataReceiver} components directly map onto the respective entities of the MTV. In particular,  \texttt{DataProvider} and \texttt{DataReceiver} are modeled as sub-components of the \texttt{PhysicalTwin}, enabling its interaction with \texttt{P2DAdapter} and \texttt{D2PAdapter} via suitable ports representing the flow of information.  

\vspace{1em}
\noindent \textbf{P2DAdapter and D2PAdapter.} 

\noindent \texttt{P2DAdapter} and \texttt{D2PAdapter} components directly map onto the respective entities of the MTV. The former receives data from the \texttt{DataProvider} and transforms and prepares them before sending them to the specific components devoted to data management. The latter, dually, transforms information (typically commands) coming from the components responsible for feedback generation into a format compatible with the physical infrastructure before sending them to the \texttt{DataReceiver}. 

\vspace{1em}
\begin{sloppypar}
\noindent \textbf{DataManager, DataProcessor, StorageManager and SharedStorage.}
\end{sloppypar}

\noindent 
Raw data provided by \texttt{P2DAdapter} are cleaned, filtered, and organized into a standardized format that aligns with the reference data model by the \texttt{DataProcessor} component. These standardized data are used by the \texttt{ShadowManager} to create and organize digital shadows based on predefined types, and is fundamental to enable other critical operations such as feeding digital models and monitoring system behavior. By combining the historical data traces with simulation results, the system can achieve comprehensive analysis and predictive capabilities. 

To support all above mentioned operations, the data are organized in a \texttt{SharedStorage} component: it acts as a passive component, serving as a central repository for securely storing processed data, digital shadows and outcomes generated by DTCs. Clearly, due to the distributed nature of a DT, the shared storage is not constrained to a single physical repository but may encompass multiple storage systems tailored to specific requirements. The organization, storage, and retrieval of data to/from the \texttt{SharedStorage} component is managed by the \texttt{StorageManager} component, which exposes a \textit{CRUDop} interface  offering create, read, update, and delete operations. 
Functionalities of both \texttt{DataProcessor} and \texttt{StorageManager} are encapsulated within the \texttt{DataManager} component, which exposes the \textit{CRUDop} interface provided by \texttt{StorageManager} directly to other DT components.

\vspace{1em}
\noindent \textbf{ShadowManager, ModelManager, ModelEngine, Simulator and TwinManager.} 

\noindent \texttt{ShadowManager} component directly maps onto the respective entity of the MTV, and is responsible for managing shadow instances. Moreover, it maps onto \texttt{DigitalShadow} entity and, in part, onto \texttt{DigitalRepresentation} entity, from which both \texttt{DigitalShadow} and \texttt{DigitalModel} derive. As mentioned before, it accesses the data provided by the \texttt{DataProcessor} through a port attachment relationship. Alternatively, depending on software requirements, it may directly access the shared storage through the \textit{CRUDop} interface. As illustrated in Figure \ref{fig:component}, \texttt{ShadowManager} provides the \textit{getShadow} API, depicted using UML lollipop notation, enabling queries on shadows based on attributes such as type, timestamp, or name.

\texttt{ModelManager} component directly corresponds to the respective entity in the MTV and, consequently, to \texttt{Digital Representation} entity from which a model is derived. It is responsible for managing the lifecycle of digital models, including their creation, update, and configuration, ensuring consistency and accuracy across the models. \texttt{ModelEngine} component handles the execution of simulations, generating results based on modeled scenarios and providing insights into potential system behaviors under various conditions. It provides the \textit{modelExecution} interface, which is utilized by \texttt{ModelManager} to execute digital models. 

Together, they constitute \texttt{Simulator} component, which combines the virtualization of real-world systems through digital models with their execution in different scenarios. Moreover,  \texttt{Simulator} offers the \textit{getSimState} and \textit{scenarioSim} APIs for simulating scenarios, enabling actions on digital models, adjusting simulation parameters, and retrieving the current simulation state. These functionalities are utilized by \texttt{TwinManager} (that directly maps onto the respective entity of the MTV) for comprehensive system orchestration to coordinate the interactions with other service-related components in order to facilitate seamless integration and efficient execution of Digital Twin services.

\vspace{1em}
\noindent \textbf{StateMonitor.} 

\noindent \texttt{StateMonitor} component enables to concretely realize the monitoring capabilities of a DT and can thus be mapped onto \texttt{ServiceManager} entity. In particular, it offers the \textit{getState} interface to \texttt{TwinManager} to retrieve the current state of the Physical Twin. The \texttt{StateMonitor} component, in particular, is responsible for collecting both real-world and simulated states, calculating the current state of the Physical Twin, storing in the shared storage for further analysis, and potentially presenting it to users via human-machine interfaces.

\begin{boxC}
\footnotesize
\texttt{StateMonitor} is fully supported by Azure Digital Twins and FIWARE. Azure Stream Analytics handles state data collection and analysis, while FIWARE leverages Cosmos for big data analysis. Eclipse Ditto provides partial support with event-handling mechanisms for processing state changes. 

\vspace{1em}
\textbf{Exemplars of component usage in a real scenario}: In the context of urban mobility and traffic flow management,  \texttt{StateMonitor} continuously collects data from traffic sensors, such as vehicle counts at intersections, average speeds on roads, and real-time GPS data from public transports. These data are combined with simulated traffic states generated by \texttt{Simulator}. For instance,  \texttt{StateMonitor} may calculate the current state as: \textit{``Intersection A has a traffic density of 80\% with an average vehicle speed of 15 km/h''}.  

\texttt{TwinManager} accesses the traffic state through the \textit{getState} interface to coordinate actions. For instance, based on the retrieved traffic data, \texttt{TwinManager} might initiate a prediction of alternative traffic flow scenarios to optimize road usage and alleviate congestion.
\end{boxC}

\noindent \textbf{DeviationDetector, Predictor, and Analyzer.} 

\noindent \texttt{DeviationDetector}, \texttt{Predictor} and \texttt{Analyzer} components specialize \texttt{ServiceManager} entity for what concerns the identification of anomalies or deviations, by comparing real or predicted states with expected states. This functionality allows the system to promptly detect discrepancies. In particular, \texttt{Predictor} forecasts future system states using current and historical data, supporting proactive decision-making. 
  
\texttt{DeviationDetector} identifies deviations based on the predicted state and the current state. \texttt{Analyzer} integrates the functionalities of \texttt{DeviationDetector} and \texttt{Predictor}, enabling analysis of real and simulated states and exposing the \textit{prediction} API offered by the \texttt{Predictor} component to \texttt{TwinManager}, which uses it to trigger forecasting mechanisms.

\begin{boxC}
\footnotesize
\texttt{DeviationDetector}, \texttt{Predictor}, and \texttt{Analyzer} are fully supported by Azure Digital Twins, partially by FIWARE, and minimally by Ditto. Azure DTs provides robust tools like the Multivariate Detection Toolkit, Azure Stream Analytics, and Azure Machine Learning for detection, prediction, and analysis. FIWARE supports these functionalities with Cosmos and Perseo for rule-based detection and forecasting. In contrast, Ditto offers limited data analytics capabilities and lacks native support for deviation detection or prediction.
\end{boxC}

\noindent \textbf{ScenarioGenerator, SolutionFinder, and Planner.} 

\noindent \texttt{ScenarioGenerator}, \texttt{SolutionFinder}, and \texttt{Planner} components specialize \texttt{ServiceManager} entity for what concerns the generation of diverse scenarios to simulate and evaluate the system's behavior under varying conditions, facilitating the exploration of alternative strategies. In particular, \texttt{Solution Finder} determines the optimal set of actions required to restore the system to its desired state when deviations or anomalies are identified. Building on this,  \texttt{Planner} leverages the solutions identified by \texttt{SolutionFinder} and analyzes the simulation results from scenarios created by  \texttt{ScenarioGenerator}. This enables the \texttt{Planner} to devise and execute effective corrective actions, ensuring the system's resilience and stability. 

As shown in Fig. \ref{fig:component}, \texttt{SolutionFinder} is connected to \texttt{DeviationDetector} through a port attachment relationship to receive real or predicted deviations. Additionally, \texttt{Scenario Generator} offers the \textit{genScenario} interface used by \texttt{Solution Finder}. Finally, \texttt{Planner} interacts with \texttt{TwinManager} through the \textit{newScenarioSim} API. This allows the \texttt{Planner} to simulate alternative scenarios by leveraging the \texttt{TwinManager}'s access to underlying simulators. 

\begin{boxC}
\footnotesize
 \texttt{SolutionFinder}, \texttt{ScenarioGenerator}, and \texttt{Planner} lack direct support from the selected technologies. Azure Digital Twins, Eclipse Ditto, and FIWARE do not provide dedicated tools for scenario generation, solution finding, or planning functionalities. These components would require custom implementations or integrations with external simulation and planning frameworks to achieve their intended purpose. 
\end{boxC}

\noindent \textbf{FeedbackExecutor.}

\noindent \texttt{FeedbackExecutor} component maps onto the \texttt{FeedbackProvider} entity of the MTV and plays a dual role: when \texttt{Deviation Detector} identifies a deviation but no corrective plan is provided by the DT, \texttt{FeedbackExecutor} generates alerts or warnings for the physical system. Conversely, when \texttt{Planner} provides a plan, \texttt{FeedbackExecutor} translates this plan into actionable instructions executable by the physical system. Then, as anticipated, \texttt{D2PAdapter} ensures that these instructions are transformed into a format compatible with the physical infrastructure.

\begin{boxC}
\footnotesize
\texttt{FeedbackExecutor} is not natively supported by the selected technologies, as its implementation depends on custom solutions tailored to the physical system and the specific objectives of the DT.

\vspace{1em}
\textbf{Exemplars of component usage in a real scenario}: If a traffic congestion is detected on Main Street without a plan, the \texttt{FeedbackExecutor} may generate an alert: \textit{``High congestion detected on Main Street; notify drivers to avoid the area''}. When a plan is available, such as \textit{``Divert vehicles from Main Street to Elm Avenue and extend green light duration on Elm Avenue by 20 seconds''},  the \texttt{FeedbackExecutor} translates it into commands for traffic control systems, e.g traffic lights and dynamic road signs.


\end{boxC}

\subsection{Traceability Twin View}
\label{sec:traceview}

\begin{table*}[!ht]
\centering
\caption{Traceability Twin View: Matrix Diagram.}
\label{tab:matrix}
\resizebox{\textwidth}{!}{%
\begin{tabular}{|l|>{\hspace{0pt}}m{0.075\linewidth}|>{\hspace{0pt}}m{0.075\linewidth}|>{\hspace{0pt}}m{0.075\linewidth}|>{\hspace{0pt}}m{0.075\linewidth}|>{\hspace{0pt}}m{0.075\linewidth}|>{\hspace{0pt}}m{0.075\linewidth}|>{\hspace{0pt}}m{0.075\linewidth}|>{\hspace{0pt}}m{0.075\linewidth}|>{\hspace{0pt}}m{0.075\linewidth}|>{\hspace{0pt}}m{0.075\linewidth}|>{\hspace{0pt}}m{0.075\linewidth}|>{\hspace{0pt}}m{0.075\linewidth}|>{\hspace{0pt}}m{0.075\linewidth}|>{\hspace{0pt}}m{0.075\linewidth}|>{\hspace{0pt}}m{0.075\linewidth}|>{\hspace{0pt}}m{0.075\linewidth}|}
\hline
\textbf{ \begin{tabular}[c]{@{}l@{}}DTE/DTC \\ Name\end{tabular}} & \begin{tabular}[c]{@{}l@{}}Physical \\ Twin\end{tabular} & \begin{tabular}[c]{@{}l@{}}Data \\ Provider \end{tabular} & \begin{tabular}[c]{@{}l@{}}Data \\ Receiver\end{tabular} & Adapter & \begin{tabular}[c]{@{}l@{}}P2D \\ Adapter\end{tabular} &  \begin{tabular}[c]{@{}l@{}}D2P \\ Adapter\end{tabular} &  \begin{tabular}[c]{@{}l@{}}Digital \\ Repres.\end{tabular} &  \begin{tabular}[c]{@{}l@{}}Digital \\ Shadow\end{tabular} & \begin{tabular}[c]{@{}l@{}}Shadow \\ Manager\end{tabular} &  \begin{tabular}[c]{@{}l@{}}Digital \\ Model\end{tabular} &  \begin{tabular}[c]{@{}l@{}}Model \\ Manager\end{tabular} &  \begin{tabular}[c]{@{}l@{}}Twin \\ Manager\end{tabular} &  \begin{tabular}[c]{@{}l@{}}Service \\ Manager\end{tabular} &  \begin{tabular}[c]{@{}l@{}}Feedback \\ Provider\end{tabular} &  \begin{tabular}[c]{@{}l@{}}Data \\ Manager\end{tabular} &  \begin{tabular}[c]{@{}l@{}}Data \\ Model\end{tabular} \\ \hline
PhysicalTwin & \checkmark &  &  &  &  &  &  &  &  &  &  &  &  &  &  &  \\ \hline
DataProvider &  & \checkmark &  &  &  &  &  &  &  &  &  &  &  &  &  &  \\ \hline
DataReceiver &  &  & \checkmark &  &  &  &  &  &  &  &  &  &  &  &  &  \\ \hline
P2DAdapter &  &  &  & \checkmark & \checkmark &  &  &  &  &  &  &  &  &  &  &  \\ \hline
D2PAdapter &  &  &  & \checkmark &  & \checkmark &  &  &  &  &  &  &  &  &  &  \\ \hline
DataProcessor &  &  &  &  &  &  &  &  &  &  &  &  &  &  & \checkmark & \checkmark \\ \hline
StorageManager &  &  &  &  &  &  &  &  &  &  &  &  &  &  & \checkmark &  \\ \hline
DataManager &  &  &  &  &  &  &  &  &  &  &  &  &  &  & \checkmark &  \\ \hline
SharedStorage &  &  &  &  &  &  &  &  &  &  &  &  &  &  & \checkmark & \checkmark \\ \hline
ShadowManager &  &  &  &  &  &  & \checkmark & \checkmark & \checkmark &  &  &  &  &  &  &  \\ \hline
ModelManager &  &  &  &  &  &  & \checkmark &  &  &  & \checkmark &  &  &  &  &  \\ \hline
ModelEngine &  &  &  &  &  &  &  &  &  & \checkmark & \checkmark &  &  &  &  &  \\ \hline
Simulator &  &  &  &  &  &  & \checkmark &  &  & \checkmark & \checkmark &  &  &  &  &  \\ \hline
TwinManager &  &  &  &  &  &  &  &  &  &  &  & \checkmark &  &  &  &  \\ \hline
StateMonitor &  &  &  &  &  &  &  &  &  &  &  &  & \checkmark &  &  &  \\ \hline
DeviationDet. &  &  &  &  &  &  &  &  &  &  &  &  & \checkmark &  &  &  \\ \hline
Predictor &  &  &  &  &  &  &  &  &  &  &  &  & \checkmark &  &  &  \\ \hline
Analyzer &  &  &  &  &  &  &  &  &  &  &  &  & \checkmark &  &  &  \\ \hline
SolutionFinder &  &  &  &  &  &  &  &  &  &  &  &  & \checkmark &  &  &  \\ \hline
ScenarioGen. &  &  &  &  &  &  &  &  &  &  &  &  & \checkmark &  &  &  \\ \hline
Planner &  &  &  &  &  &  &  &  &  &  &  &  & \checkmark &  &  &  \\ \hline
FeedbackExec. &  &  &  &  &  &  &  &  &  &  &  &  &  & \checkmark &  &  \\ \hline
\end{tabular}%
}
\end{table*}

The \textit{Traceability Twin View} serves as the bridge between the module and component views of the proposed TwinArch. As outlined by the SEI \cite{Book_DocumentingSA}, the architectural elements across different views must be interrelated to maintain coherence across abstraction levels, ensuring that high-level domain concepts are effectively aligned with the software components that implement them.

The separation of concerns inherent to this approach is pivotal. The Module Twin View focuses on defining the system's objectives and functionalities, outlining what the system is designed to accomplish. In contrast, the Component Twin View delves into the structural and operational aspects, detailing how these objectives are realized. By connecting these two perspectives, the Traceability Twin View ensures consistency, enhances clarity, and fosters alignment across all levels of the architecture, providing a unified understanding of the system's design and implementation.

The \textit{Matrix Diagram} in Table \ref{tab:matrix} illustrates the relationships between MTV entities and their corresponding CTV components. As discussed in the previous subsections, some mappings are straightforward, such as \texttt{PhysicalTwin}, \texttt{DataProvider}, and \texttt{DataReceiver}. These elements primarily relate to the physical system, and their further refinement is domain-dependent.  In other cases, domain entities are mapped to multiple components. For instance, \texttt{DataManager} and \texttt{DataModel} entities correspond to \texttt{DataProcessor}, \texttt{Storage Manager}, and \texttt{SharedStorage} components. This mapping is driven by the adoption of the shared-data architectural style in CTV, designed to handle the foundational role of data in a Digital Twin system.

While mappings for digital shadows and models are clear, \texttt{ServiceManager} exemplifies a more nuanced case. This class abstracts the idea of services in the module view, while in the CTV the concept of DT services is refined through the introduction of the following components: \texttt{StateMonitor}, \texttt{DeviationDetector}, \texttt{Predictor}, \texttt{Analyzer}, \texttt{SolutionFinder}, \texttt{ScenarioGenerator}, and \texttt{Planner}. These components reflect the Digital Twin system's service-oriented capabilities, supporting monitoring, deviation detection, prediction, data analysis, and planning. 

\subsection{Dynamic Twin View}
\label{sec:dynview}
The \textit{Dynamic Twin View} captures the runtime interactions between the architectural elements of the DT to achieve specific functionalities. This perspective is illustrated through two use cases. The first use case focuses on \textit{monitoring}, which entails the continuous tracking and analysis of the physical system's state by combining simulations of digital models with real-world data. The second use case addresses the \textit{prediction}, leveraging the DT's simulation and analytical capabilities to forecast future states or behaviors of the physical system, including identifying potential issues, evaluating alternative scenarios, and recommending corrective actions. 
Both use cases are represented using \textit{UML Sequence Diagrams}. The monitoring use case, which employs DT domain entities, is detailed in Section~\ref{sec:monitoring}, while the prediction use case, utilizing DT components, is explained in Section~\ref{sec:prediction}. This distinction emphasizes the modeling of dynamic interactions at varying levels of abstraction (entities and components) to effectively address diverse functionalities.

\subsubsection{Use Case: Monitoring Service}
\label{sec:monitoring}
\begin{figure*}[!ht]
\centerline{\includegraphics[width=0.7\textwidth]{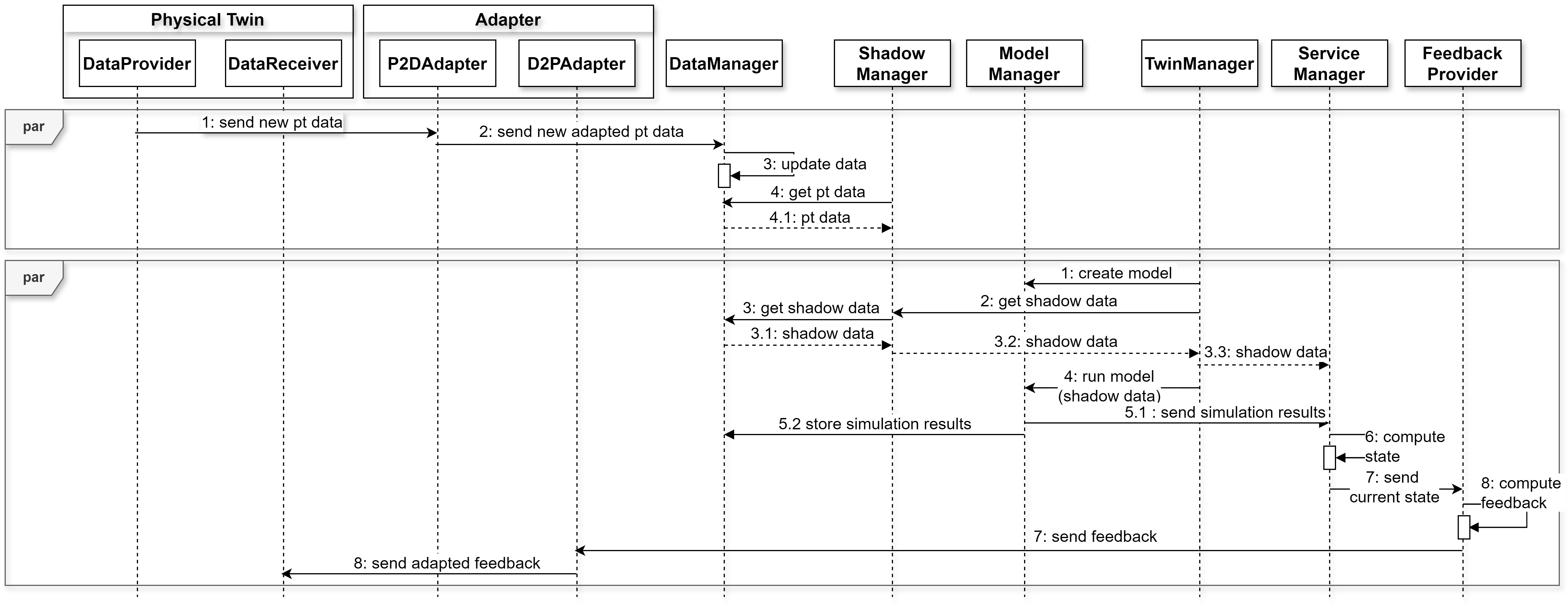}}
\caption{Dynamic Twin View: UML Sequence Diagram of Monitoring Use Case with domain entities.}
\label{fig:sdmonitoring}
\end{figure*}

Figure \ref{fig:sdmonitoring} illustrates the \textit{UML Sequence Diagram} of monitoring use case. The process begins with\texttt{DataProvider} transmitting data collected from the physical twin to the DT system. These data are processed by \texttt{P2DAdapter}, which converts it into a format compatible with the DT. The adapted data are subsequently forwarded to the \texttt{DataManager} to efficiently store and manage them. \texttt{ShadowManager} retrieves these data from \texttt{DataManager} (or directly via \texttt{P2DAdapter} to reduce latency) to update the relevant digital shadows. These shadows represent real-world states grouped by shadow types (e.g., traffic packets or other domain-specific data types).

As \texttt{PhysicalTwin} continues to send data, \texttt{ModelManager} is engaged to create or update digital models, used by \texttt{Twin Manager} to execute simulations by feeding the models with the real-world data. The results of these simulations are stored in the system, providing enhanced insights into system behavior. In the context of monitoring, \texttt{ServiceManager} computes the state of the physical system by combining the simulation results with the shadow data. This computed state is then delivered to \texttt{FeedbackProvider}, which generates feedback messages. For instance, a simple message such as \textit{“system ok”} may be generated, completing the monitoring loop. This feedback enables actionable responses, allowing the physical system to adjust as needed.

\subsubsection{Use Case: Prediction Service}
\label{sec:prediction}
\begin{figure*}[!h]
\centerline{\includegraphics[width=0.65\textwidth]{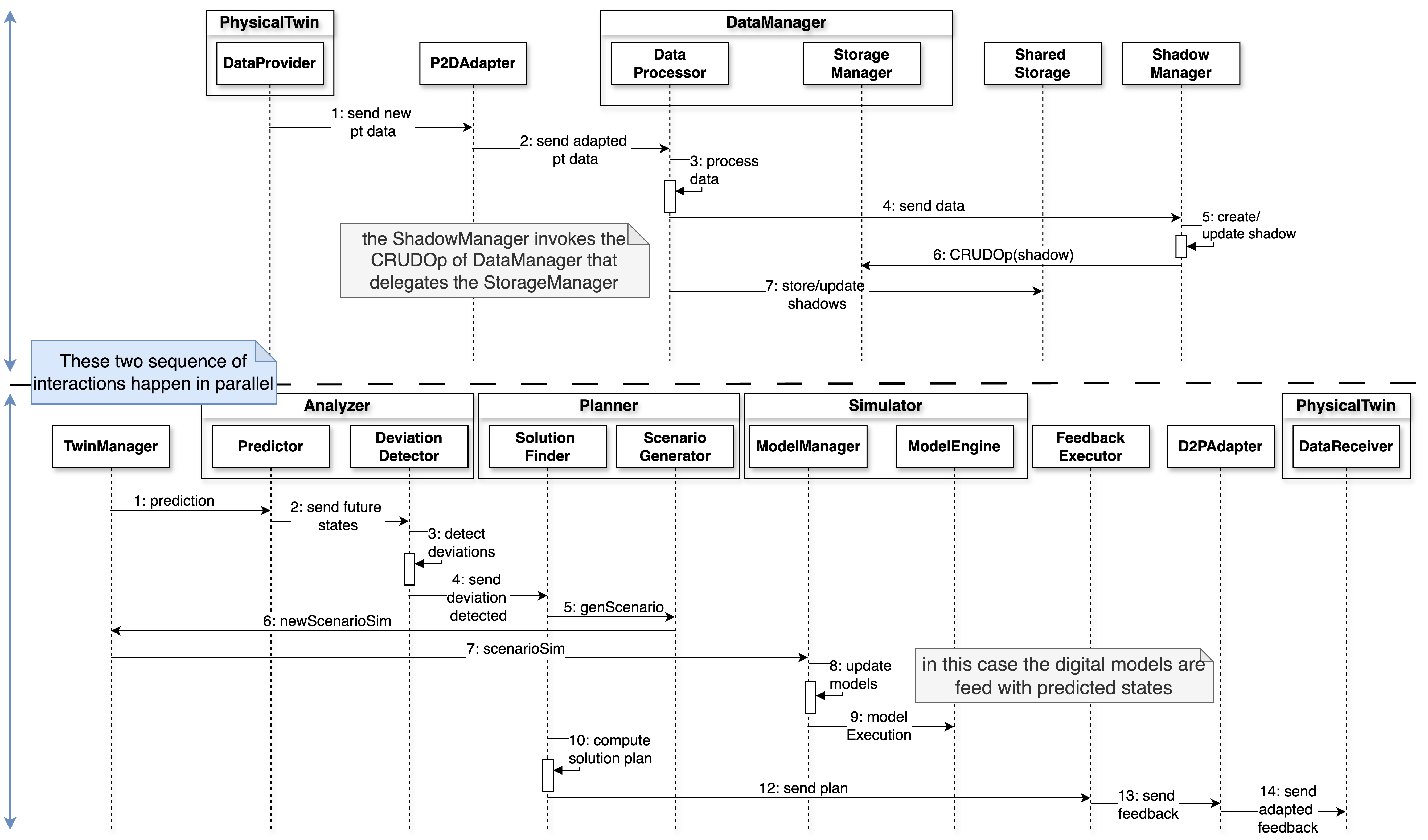}}
\caption{Dynamic Twin View: UML Sequence Diagram of Prediction Use Case with components.}
\label{fig:sdprediction}
\end{figure*}

Figure \ref{fig:sdprediction} illustrates the \textit{UML Sequence Diagram} for the prediction use case, in which the DT forecasts future states of the physical system, identifies potential failures, and determines optimal actions to prevent them.
The process begins with \texttt{TwinManager} initiating a prediction request, which prompts \texttt{Predictor} component to forecast future states of the system. 

\texttt{Predictor} transmits the computed future states to the \texttt{DeviationDetector}, responsible for identifying potential deviations by comparing the predicted states against predefined thresholds or expected values. If a deviation is detected, \texttt{DeviationDetector} notifies \texttt{SolutionFinder} to initiate the deviation-handling process. \texttt{SolutionFinder} collaborates with \texttt{ScenarioGenerator} to explore and evaluate alternative scenarios for resolving the detected deviation.

When \texttt{SolutionFinder} identifies the possible solutions to address the problem, it triggers \texttt{ScenarioGenerator} to create detailed simulations of these solutions. Therefore,  the execution of these new scenarios are triggered by interacting with \texttt{Simulator} through \texttt{TwinManager}. The manager facilitates the communication to update or create digital models via \texttt{ModelManager}, which subsequently triggers models execution by invoking \texttt{ModelEngine}.


These simulations provide critical insights into potential outcomes, enabling \texttt{SolutionFinder} to compute the optimal solution plan. Once the solution plan is finalized, it is transmitted to \texttt{FeedbackExecutor}, which ensures that the plan is translated into actionable feedback for \texttt{PhysicalTwin}. The feedback is formatted appropriately using  \texttt{D2PAdapter} and sent to \texttt{DataReceiver}. Finally, the physical system applies this feedback to adjust its operations, thereby closing the prediction loop.

\section{Online Survey}
\label{sec:survey}
This Section presents the outcomes of the validation conducted with industry and academic experts through the online questionnaire\footnote{Available at: \url{https://alessandrasomma28.github.io/twinarch/validation.html}}. Detailed results are available in the replication package. 

The final validation aims to evaluate whether the proposed TwinArch is complete, useful, and usable, ensuring the reference architecture's effectiveness in facilitating the design and implementation of DT systems across diverse domains.
To achieve this goal, three hypotheses were formulated for each architectural view:
\begin{enumerate}[topsep=0pt,itemsep=-1ex,partopsep=1ex,parsep=1ex, label=\textit{H\arabic*.}, leftmargin=10mm]
    \item The Architectural View adequately represents all necessary elements (\textit{completeness}).
    \item The Architectural View provides practical value for DT design and development (\textit{usefulness}). 
    \item The Architectural View is intuitive and easy to apply in practice (\textit{perceived usability}). 
\end{enumerate}
These attributes are crucial for capturing practitioners' evaluations of the core dimensions that determine the impact of the proposed architecture \cite{usefulness,usability}. \textbf{\textit{Completeness}} ensures that the architectural view includes all essential elements, avoiding any gaps that could compromise system functionality. \textbf{\textit{Usefulness}} confirms that the architecture delivers practical value by effectively supporting and guiding the design and development of Digital Twin systems. Lastly, \textbf{\textit{perceived usability}} represents how easy to use the given artifacts are perceived by the practitioners. 


\begin{figure*}[ht!]
\centerline{\includegraphics[width=0.8\textwidth]{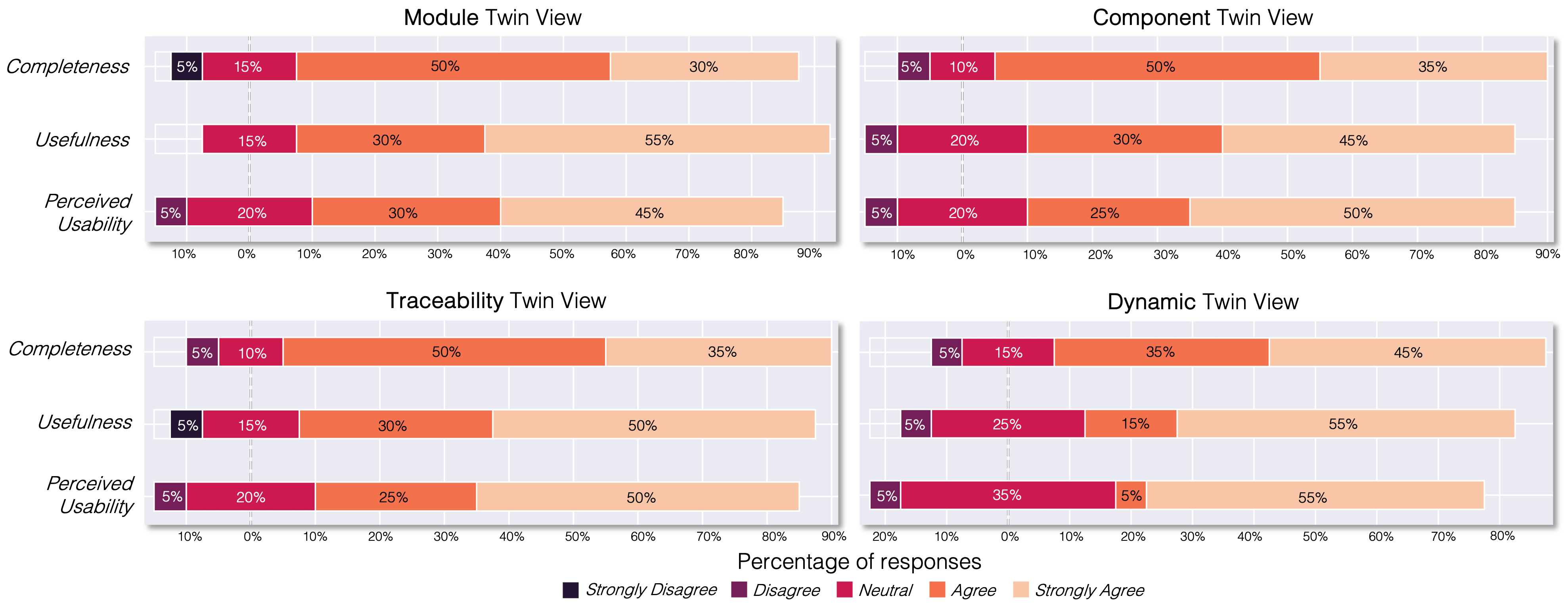}}
\caption{Likert plot for questionnaire answers.}
\label{fig:likertplot}
\end{figure*}

\subsection{Questionnaire results}
\label{sec:results}
Figure \ref{fig:likertplot} presents a summary of the questionnaire responses, capturing the perceptions of all 20 respondents regarding the completeness, usefulness, and perceived usability of TwinArch.

As shown in the Likert plots, TwinArch is widely considered complete, with a significant majority affirming the completeness of its module, component, traceability, and dynamic views (35–50\% agree, 30–45\% strongly agree with hypothesis \textit{H1}). Many respondents emphasized that TwinArch ``\textit{aims at giving a very complete architectural description of how a Digital Twin should be structured}'' and considered it particularly important because ``\textit{at the moment a unified definition/description of a Digital Twin system is not present}''. The 15\% of respondents provided a neutral response, which, as justified in open-ended questions, stemmed from the perception that completeness is subjective and dependent on the specific goals of the Digital Twin being developed. However, a small percentage (5\%) disagreed or strongly disagreed with the completeness of TwinArch, citing the lack of architectural elements for user interfaces responsible for external interfacing and enhancement of data management aspects.

TwinArch is positively evaluated in terms of usefulness, with 45–55\% of participant strongly agreeing with hypothesis \textit{H2}. Practitioners highlighted that it is ``\textit{a well-established way to document architecture and design a DT}'', particularly facilitating discussions among diverse groups such as stakeholders, system engineers, and developers. The selection of the UML language received positive feedback, and the traceability between views was also appreciated for its ability to map elements effectively, with one respondent stating that ``\textit{there is coherence among the various views}''. Respondents acknowledged that TwinArch ``\textit{gives some useful guidelines that can be followed by developers and researchers in the Digital Twin domain}'' and provides ``\textit{a clear and practical reference guideline, making it valuable for researchers, companies, and developers aiming to implement digital twin solutions}''. These comments emphasize the practical usefulness of TwinArch as a guideline for developing Digital Twins across various domains.

Regarding perceived usability, the responses included more neutral and disagreeing opinions compared to the other attributes. While practitioners recognized that TwinArch ``\textit{can be part of a DT standard}'', the ease of use aspect generated some disagreement. This was primarily attributed to the domain independence of the proposal, as respondents highlighted that ``\textit{it would have to be declined in the future on the specific applications to make it operational}''. Some practitioners noted the ``\textit{lack of practical examples}'' and ``\textit{limited stakeholder engagement}'' as factors hindering usability. Additionally, the ease of use was perceived as challenging when trying to extend TwinArch adoption to stakeholders with ``\textit{limited technical background or familiarity with software architecture}''. 

\subsubsection{Statistical Analysis}
\begin{figure*}[ht!]
\centerline{\includegraphics[width=0.8\textwidth]{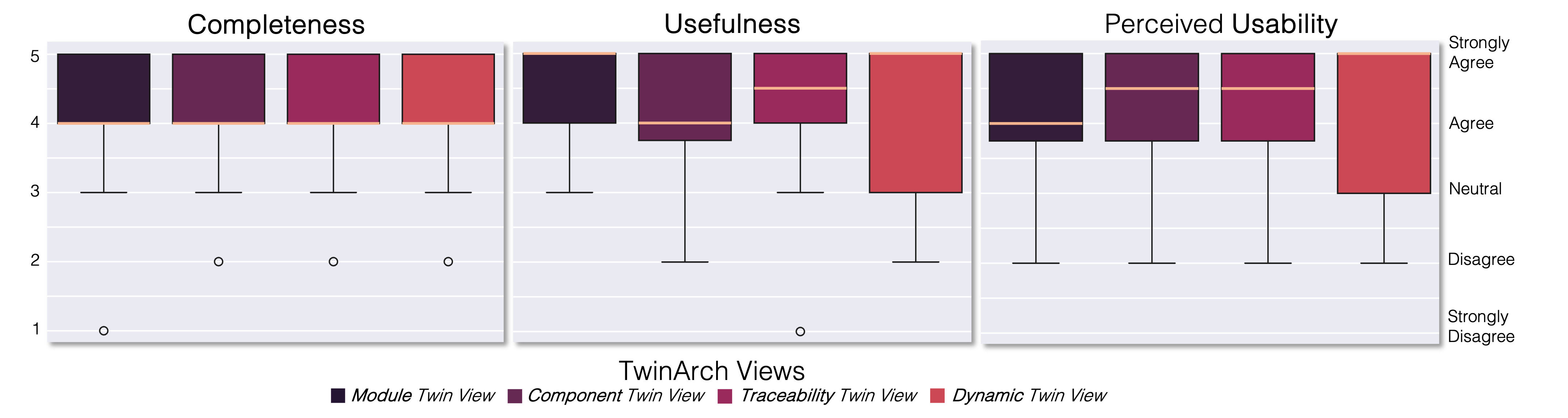}}
\caption{Box plots for questionnaire answers.}
\label{fig:boxplot}
\end{figure*}
In addition to the Likert plot analysis, we performed a detailed statistical analysis of the responses. First, we used boxplots to visualize where the hypotheses \textit{H1}, \textit{H2}, and \textit{H3} hold for each view. Considering that responses follow a Likert scale where Strongly Disagree $= 1$, Disagree $= 2$, Neutral $= 3$, Agree $= 4$, and Strongly Agree $= 5$, the resulting boxplots are shown in Figure \ref{fig:boxplot}. These plots compare the three attributes—completeness, usefulness, and perceived usability—across the different views. As rendered in Figure \ref{fig:boxplot}, the medians for all attributes consistently fall within the Agree and Strongly Agree categories across all views, reflecting an overall positive perception of TwinArch. This indicates that respondents perceive the synthesized TwinArch as complete, useful, and usable, consistent with the findings from the Likert plot analysis.

Additionally, we conducted a deeper statistical analysis to examine the completeness, usefulness, and perceived usability of the proposed reference architecture. To achieve this, we transformed the responses as follows: Strongly Disagree and Disagree (1 and 2) were mapped to -1, Neutral (3) to 0, and Agree and Strongly Agree (4 and 5) to 1. This transformation creates a symmetric scale centered at zero, ensuring that negative and positive responses are equally distanced. By summing the transformed responses for each respondent, we determine whether the aggregate score supports the hypotheses ($sum > 0$) or contradicts them ($sum < 0$).

To determine the appropriate statistical test, we first assessed data normality using the Shapiro-Wilk test, which indicated that data did not follow a normal distribution. Consequently, we employed the non-parametric one-sample Wilcoxon test to evaluate statistical and practical significance. 

\begin{figure}[h!]
\centerline{\includegraphics[width=0.8\columnwidth]{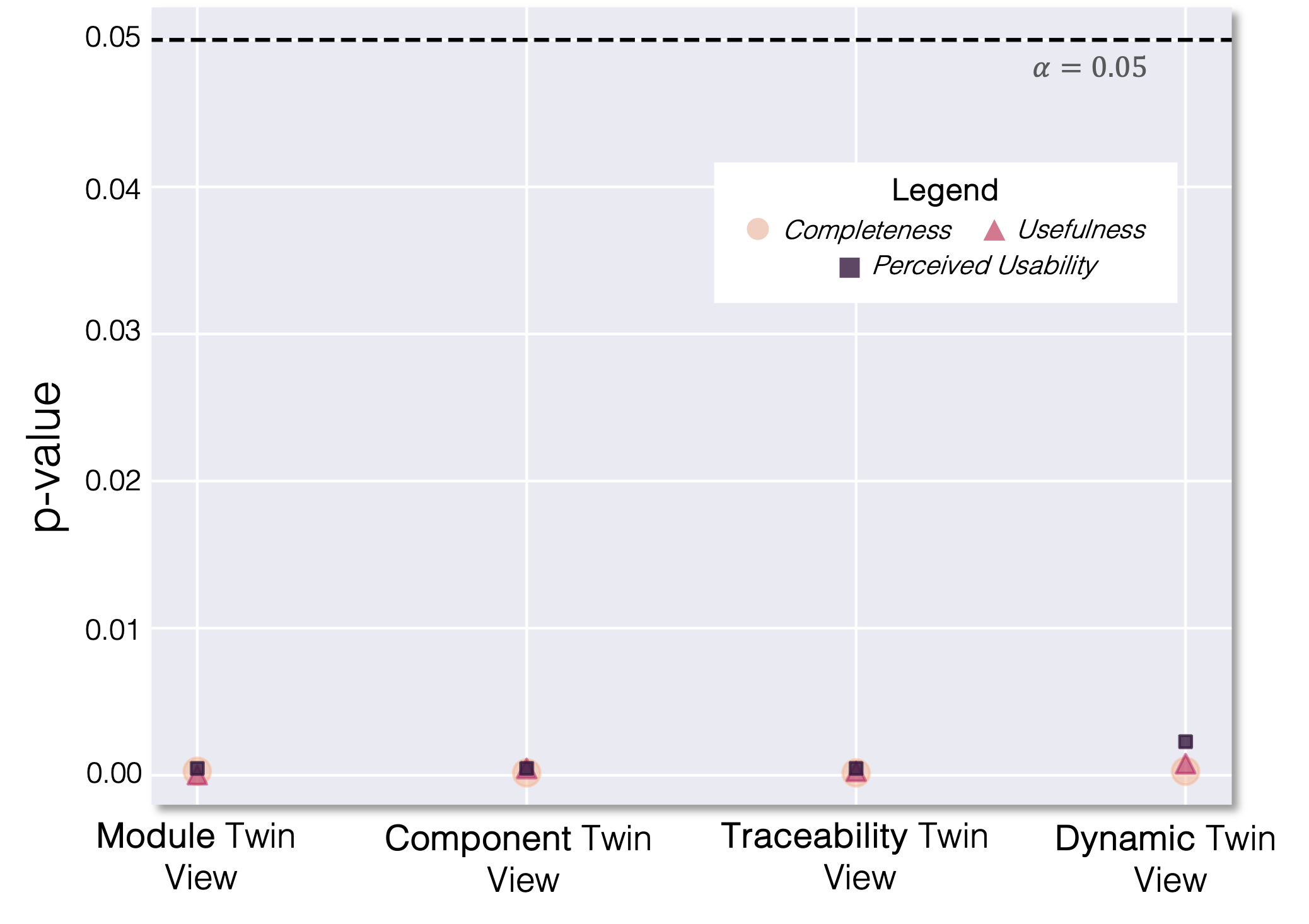}}
\caption{Wilcoxon one-sample test for statistical significance (hypothesis: $\mu \geq 0$).}
\label{fig:wilxplot}
\end{figure}

\noindent \textbf{Statistical Significance.}  
We tested our hypotheses for statistical significance using the one-sample Wilcoxon signed-rank test, which determines whether the aggregate responses for each view and dimension are significantly greater than 0 at a 5\% significance level. The results are illustrated in Figure~\ref{fig:wilxplot}. 

We note that all p-values for each view and each dimension are below the $\alpha=0.05$ threshold, indicating statistically significant results. Additionally, the majority of p-values are very close to 0, confirming that most respondents rated the TwinArch attributes well above the neutral value. The results validate that the correlation between the TwinArch views and the quality attributes (completeness, usefulness, and perceived usability) is statistically significant. This finding further underscores the relevance and applicability of TwinArch as a reference architecture for DT systems.

\begin{figure}[h!]
\centerline{\includegraphics[width=0.9\columnwidth]{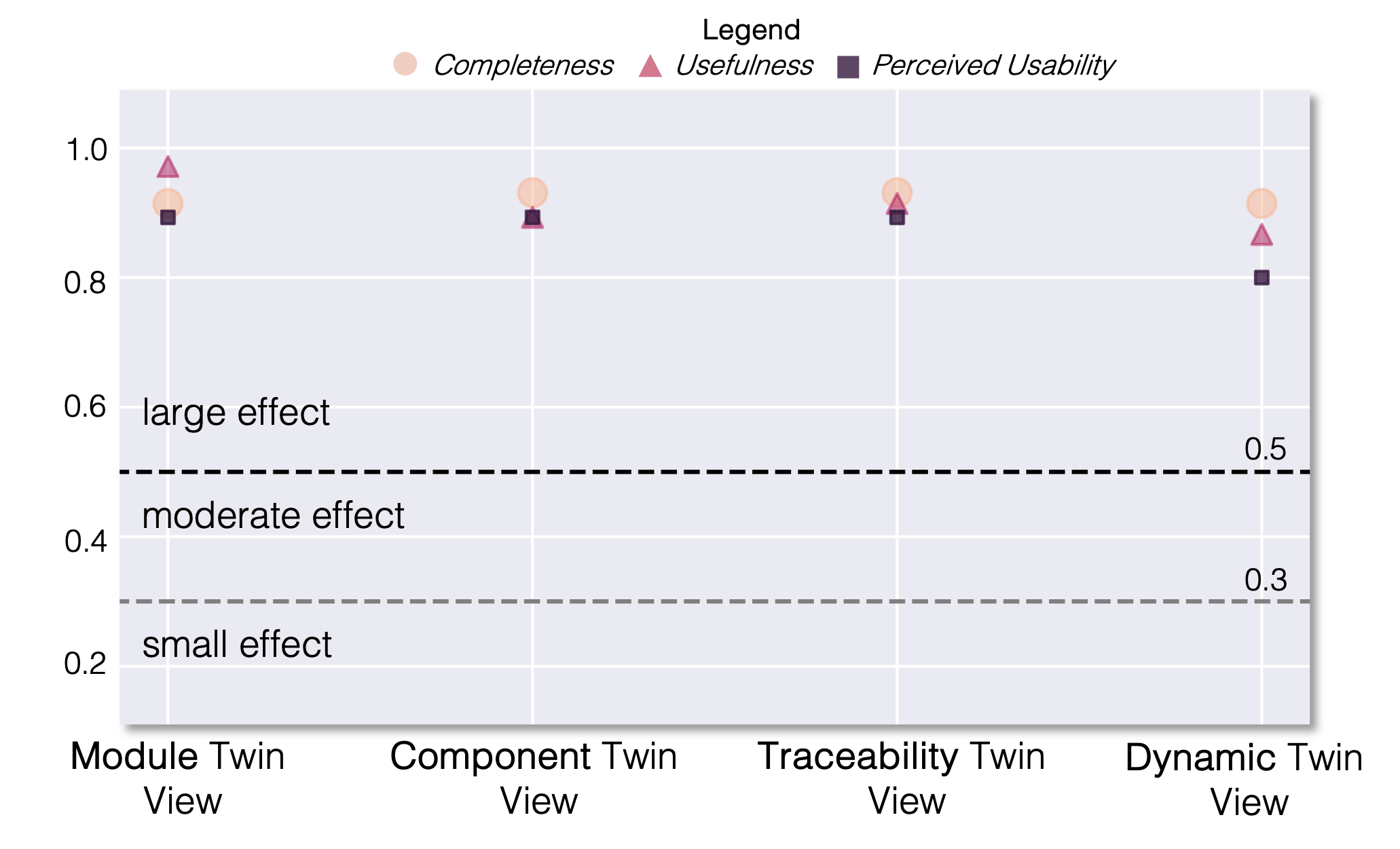}}
\caption{Effect size for one sample Wilcoxon test for practical significance.}
\label{fig:practplot}
\end{figure}

\noindent \textbf{Practical Significance.} We tested our hypotheses for practical significance calculating the Cohen's measure of effect size $r$ derived from the one-sample Wilcoxon signed-rank test. The test evaluates whether the effect size is greater than zero for the sum of all respondents’ answers across combinations of views and attributes. The resulting plot is shown in Figure \ref{fig:practplot}.

According to Cohen J. \cite{cohen}, the thresholds for effect size are defined as small effect $< 0.3$, moderate effect $>0.3$ and $<0.5$, and large effect $>0.5$. From Fig. \ref{fig:practplot}, it is clear that all effect sizes exceed the large effect's threshold across all views and attributes. This demonstrates the practical significance of TwinArch's completeness, usefulness, and perceived usability.

More in detail, completeness and usefulness consistently show higher effect sizes across all views compared to perceived usability, suggesting that participants found the former attributes to be particularly impactful. This difference may be attributed to the limited availability of practical Digital Twin examples, which would allow for a more thorough evaluation of the ease-of-use attribute. Nevertheless, while slightly lower, perceived usability still falls well within the range of a large effect. These findings confirm that TwinArch not only achieves statistical significance but also demonstrates meaningful practical utility across its views and attributes.

\section{Discussion}
\label{sec:discussion}
This Section discusses the limitations identified in the selected studies on Digital Twin architectures, and how TwinArch addresses these challenges. Additionally, we examine the mapping between TwinArch's elements and the ISO 23247 functional entities to assess its alignment with existing relevant standards, identifying both gaps and enhancements introduced by TwinArch.

\noindent \textit{\textbf{Lack of multi-view documentation.}} As already pointed out in Sections \ref{sec:introduction} and \ref{sec:related}, a major limitation in existing Digital Twin architectures is the lack of structured multi-view documentation \cite{M17_DTCNC23, M30_KeyComponentsDT22}. Many architectures rely on single-diagram representations that combine different abstraction levels \cite{M05_DTPort22, M09_maketwin24}, leading to confusion and misinterpretation of architectural elements. Additionally, unclear relationships between elements and the mixing of structural and dynamic aspects make it difficult to understand the runtime behavior of DT systems in use cases such as monitoring or prediction~\cite{M01_BlockchainDT24, M07_DTWaterPlatform24, M43_ModelHealthDT24, M24_DTDataspace22}.

TwinArch addresses these issues by adopting the SEI Views and Beyond method, resulting in a structured, multi-view architecture. By clearly delineating distinct architectural views, TwinArch prevents the incorrect assignment of elements to inappropriate levels of abstraction. Additionally, traceability between views enhances clarity, allowing stakeholders to seamlessly navigate between high-level conceptual overviews and detailed technical specifications. The introduction of a dedicated view for dynamic aspects and the modeling of different use cases in separate diagrams significantly improves the understanding of DT runtime behavior. 

\noindent \textit{\textbf{Lack of data-related aspects.}} A key limitation in existing DT architectures is the overemphasis on simulation functionalities, while data-related aspects remain underdeveloped despite their central role in DT systems. Many proposals focus primarily on virtual modeling \cite{M34_DTModeling21, M38_DTSixLayer20, M02_KnowledgeDT24, M16_ArchitecturePdm23}, often neglecting critical aspects such as data management, bidirectional data exchange, and data adaptation.

TwinArch addresses this gap by adopting a data-centric architectural approach, utilizing a shared-data style to decouple data producers from consumers through a shared repository for data exchange. This enhances modifiability and scalability, facilitating efficient data handling across various DT applications. Additionally, TwinArch explicitly integrates data adaptation and shadowing as core architectural components, ensuring these essential aspects are systematically incorporated rather than treated as secondary considerations.

\begin{table*}[!ht]
\centering
\caption{TwinArch's elements mapping to ISO 23247 Functional Entities.}
\label{tab:isomapping}
\resizebox{\textwidth}{!}{%
\begin{tabular}{|p{5cm}|p{6cm}|p{6cm}|p{6cm}|}
\hline
\textbf{ISO 23247 Functional Entity} & \textbf{ISO 23247 Domain} & \textbf{Module Twin View Element} & \textbf{Component Twin View Element} \\ \hline
Observable Manufacturing Elements & Observable Manufacturing Domain & PhysicalTwin & PhysicalTwin \\ \hline
Data Collecting & Data Collection and Device Control Domain & DataProvider & DataProvider \\ \hline
Data Pre-Processing & Data Collection and Device Control Domain & DataManager & DataProcessor \\ \hline
Data Translation & Cross-System Domain & Adapter, P2DAdapter, D2PAdapter & P2DAdapter, D2PAdapter \\ \hline
Controlling & Data Collection and Device Control Domain & FeedbackProvider & FeedbackExecutor \\ \hline
Actuation & Data Collection and Device Control Domain & DataReceiver & DataReceiver \\ \hline
Digital Modeling & Core Domain & DigitalRepresentation, DigitalModel & ModelEngine \\ \hline
Maintenance & Core Domain & \notatall & \notatall \\ \hline
Synchronization & Core Domain & DataProvider, DataReceiver, TwinManager & DataProvider, DataReceiver, TwinManager \\ \hline
Simulation & Core Domain & ModelManager & Simulator \\ \hline
Analytic Service & Core Domain & ServiceManager & Analyzer \\ \hline
Reporting & Core Domain & TwinManager & TwinManager, StateMonitor \\ \hline
Application Support & Core Domain & TwinManager & TwinManager \\ \hline
Interoperability Support & Core Domain & DataModel, DataManager & DataProcessor, SharedStorage \\ \hline
Access Control & Core Domain & \notatall & \notatall \\ \hline
Security Support & Cross-System Domain & \notatall & \notatall \\ \hline
User Interface & User Domain & TwinManager & TwinManager \\ \hline
\end{tabular}%
}
\end{table*}

\noindent \textit{\textbf{Domain dependence of DT architectures.}} Another challenge limiting the widespread adoption of DTs is the strong domain dependence of state-of-the-art architectures. Many existing solutions are tailored to specific industries, such as manufacturing, aerospace, and healthcare, making them less adaptable to cross-domain applications \cite{M33_SelfAdaptiveDT21, M38_DTSixLayer20, M39_DesignFramework20, M13_DTHealth23}. 

TwinArch overcomes this limitation by introducing a domain-independent reference architecture, designed for wide applicability across various DT domains. Its architectural elements are mapped onto multiple DT development platforms without being tied to a specific application domain. This allows practitioners to leverage reusable and concrete artifacts, which can be customized to meet domain-specific requirements.

\noindent \textit{\textbf{Gaps in the ISO 23247 standard.}} As noted in \cite{RW_StandardizationDTArch}, alongside its domain dependence, the ISO 23247 standard lacks comprehensive coverage of key architectural elements, particularly those related to data management. Its reliance on an entity-based reference model, with limited attention to runtime behaviors and interactions, and the absence of concrete artifacts to support practitioners, poses significant challenges for the effective development and implementation of Digital Twin solutions.

TwinArch addresses these shortcomings while aligning its architectural elements with the ISO 23247 Functional Entities. Table \ref{tab:isomapping} presents the mapping between the standard's FEs, their respective domains as defined in ISO 23247, and the corresponding Module Twin View and Component Twin View elements in TwinArch. This mapping highlights areas where TwinArch ensures compliance, providing a one-to-one alignment with key entities such as \texttt{PhysicalTwin}, \texttt{DataProvider}, and \texttt{DigitalModel}, thereby supporting the standard’s foundational definitions.

However, the table also underscores critical omissions in the ISO 23247 standard. For instance, the standard does not address data-related entities, such as digital shadowing elements which play a pivotal role in ensuring historical state management and real-time synchronization. Similarly, service-related entities like those supporting planning, scenario generation, and prediction are absent in the standard, even though these functionalities are essential for DT implementations. While basic concepts such as simulation, monitoring, and analysis are covered, the lack of advanced service-oriented elements limits the standard's applicability for more complex DT use cases. 

On the other hand, it is evident that TwinArch does not address non-functional requirements, such as security, access control, and maintenance, as these aspects fall outside the scope of the proposed architecture. By bridging the gaps in ISO 23247 and extending its capabilities, TwinArch provides a concrete and extended architecture for supporting practitioners in the design and development of DT solutions.

\section{Conclusion and Future Work}
\label{sec:conclusion}
This paper proposed a domain-independent and multi-view Digital Twin Reference Architecture, called TwinArch. TwinArch helps filling a research gap in the field since existing reference architectures and standards are tailored to a specific domain, like manufacturing, and are typically documented with a single view mixing static and dynamic aspects. 

TwinArch has been built by carefully analyzing the state of the art in the field, as well as the most common DT development platforms. The proposed reference architecture has been also validated thanks to the help of 20 experts to check its completeness, usefulness, and perceived usability. We believe that TwinArch provides practitioners with practical artifacts that can be used to design and develop new DT systems across various domains and to document existing ones.

As future work, we plan to put in practice TwinArch in various domains to better experiment its usefulness and perceived usability. We also believe that this study can help standard bodies to define a reference architecture standard.

\section{Funding}\label{sec:ack}
This work has been partially supported by the Spoke~9 ``\textit{Digital Society \& Smart Cities}'' of ICSC - Centro Nazionale di Ricerca in High Performance-Computing, Big Data and Quantum Computing, funded by the European Union - NextGenerationEU (PNRR-HPC, CUP: E63C22000980007). Moreover, this work has been partially funded by: \textit{(a)} the National Science Foundation under Grant No. 2232721; \textit{(b)} the MUR (Italy) Department of Excellence 2023 - 2027; \textit{(c)} the European HORIZON-KDT-JU research project MATISSE ``\textit{Model-based engineering of Digital Twins for early verification and validation of Industrial Systems}'', HORIZON-KDT-JU-2023-2-RIA, Prop. n.:  101140216-2, KDT232RIA\_00017; \textit{(d)} the PRIN project P2022RSW5W - RoboChor: Robot Choreography; \textit{(e)} the PRIN project 2022JKA4SL - HALO: etHical-aware AdjustabLe autOnomous systems.

\section*{Data Availability} 
\label{sec:dataaval}
The SLR process, the analysis of the selected papers, the TwinArch documentation presented through UML diagrams, the high-quality figures, as well as the online survey questionnaire and its results, are accessible at: \url{https://alessandrasomma28.github.io/twinarch/}.




\end{document}